\documentclass[11pt,english]{svjour3}
\usepackage[latin9]{inputenc}
\usepackage{amsmath}
\usepackage{amssymb}
\usepackage[numbers]{natbib}

\makeatletter

\newcommand*\LyXThinSpace{\,\hspace{0pt}}

\newenvironment{svmultproof}{\begin{proof}}{\qed\end{proof}}

\usepackage[all]{xy}
\usepackage{amsmath}
\usepackage{slashed}
\usepackage{graphicx}
\usepackage{hyperref}
\usepackage{marginnote}
\usepackage[misc,geometry]{ifsym} 

\hypersetup{colorlinks=true, linkcolor=blue, citecolor=blue, filecolor=blue, pagecolor=blue, urlcolor=red, pdfauthor={}, pdftitle={graph_reconstruction_feynman_rules}}

\makeatother

\usepackage{babel}

\begin{document}
\title{Graph Reconstruction, Functorial Feynman Rules and Superposition Principles}
\titlerunning{Graph Reconstruction, Functorial Feynman Rules and Superposition Principles}   
\author{Yuri Ximenes Martins$^{1}$   \and     Rodney Josué Biezuner$^1$}
\authorrunning{Yuri Ximenes Martins and Rodney Josué Biezuner}
\institute{\reversemarginpar
\marginnote{\Letter}Yuri Ximenes Martins \at yurixm@ufmg.br \and Rodney Josué Biezuner \at rodneyjb@ufmg.br \and \reversemarginpar
\marginnote{$^1$} Departamento de Matemática, ICEx, Universidade Federal de Minas Gerais, Av. Antônio Carlos 6627, Pampulha, CP 702, CEP 31270-901, Belo Horizonte, MG, Brazil \at }

\date{Received: date / Accepted: date}
\maketitle

\begin{abstract}
In this article functorial Feynman rules are introduced as large generalizations of physicists Feynman rules, in the sense that they can be applied to arbitrary classes of hypergraphs, possibly endowed with any kind of structure on their vertices and hyperedges. We show that the reconstruction conjecture for classes of (possibly structured) hypergraphs admit a sheaf-theoretic characterization, allowing us to consider analogous conjectures. We propose an axiomatization for the notion of superposition principle and prove that the functorial Feynman rules work as a bridge between reconstruction conjectures and superposition principles, meaning that a conjecture for a class  of hypergraphs is satisfied only if each functorial Feynman rule defined on it  induces a superposition principle. Applications in perturbative euclidean quantum field theory and graph theory are given.
\keywords{graph reconstruction \and Feynman graphs \and Feynman rules \and superposition principle}
\subclass{18D10 \and 81T18 \and 05C60 \and 81Q30} \end{abstract}

\section{Introduction}

Graphs (and their generalizations such as hypergraphs) appear in the
most different areas of mathematics and physics, generally parametrizing
definitions and constructions. For instance: in algebraic geometry
the Deligne-Mumford moduli stack of stable curves is stratified by
certain graphs \citep{Deline-Munford_1,Deligne-Munford_2}; in symplectic
topology the stable maps, which play an important role in the study
of $J$-holomorphic curves and Gromov-Witten theory, are defined by
making use of graphs \citep{stable_curves_1,stable_curves_2}; Kontsevich's
formula for deformation quantization of Poisson manifolds is parametrized
by graphs \citep{quantization_poisson_manifolds_1,quantization_poisson_manifolds_2};
the Ribbon graphs (fat graphs) are used to compute the weak homotopy
type of the geometric realization of mapping class group of surfaces
with marked points \citep{ribbon_graphs_1,ribbon_graphs_2,ribbon_graphs_3};
in perturbative quantum field theory the Feynman graphs parametrize
the possible worldlines of relativistic quantum particles \citep{TQC_1,TQC_2};
the moduli space of marked surfaces (and, therefore, stable graphs)
is also used to parametrize the worldsheet of closed strings \citep{string_1,string_2};
the recent amplituhedrons, which parametrize the scattering amplitudes,
are certain hypergraphs \citep{amplituhedron_1}.

In the abstract study of graph theory there are important conjectures,
known as \emph{reconstruction conjectures}, stating that in order
to describe a graph (belonging to a certain fixed class) it is necessary
and sufficient to describe subgraphs obtained by some deleting process.
It is known that these conjectures are true for some classes (such
as trees, regular graphs, maximal planar graphs) and false for others
(such as digraphs and infinite graphs), but the general classification
remains broadly open (see \citep{reconstruction_trees_2,reconstruction_trees_3}
for a review and an exposition).

On the other hand, in a completely different perspective, there are
the so-called \emph{superposition principles} which state that physical
properties of a physical system are totally determined by the corresponding
properties of certain subsystems. They typically occur when the physical
property in question is described by a linear partial differential
equation. For example, Maxwell's and Schr\"{o}dinger's equations
are linear, so that we have wave superposition and wave function superposition
(quantum superposition). Thinking in this way, it is natural to regard
a superposition principle as some kind of reconstruction phenomenom.

When considering systems of perturbative quantum field theory we have
both graphs (or even hypergraphs) and physical properties: Feynman
graphs and scattering amplitudes. They are related via certain rules,
known as Feynman rules \citep{TQC_1,TQC_2}. It then makes sense to
consider the reconstruction conjecture for Feynman graphs and to ask
about superposition principles for scattering amplitudes. Furthermore,
it is natural to ask if Feynman rules play some role between these
two types of reconstruction processes. In this article, our objective
is to give a positive answer, but in a much more general setup.

More precisely, we show that the physicists Feynman rules can be axiomatized
under a very general frame, being regarded as functors defined in
some category of structured hypergraphs and taking values into the
category of analytic expressions (which will define the scattering
amplitudes) of some monoidal category (which plays the role of a context
where functional analysis can be done). We call these functors \emph{Feynman
functors}. We prove that the classic Feynman rules can be extended
for structured hypergraphs and for any suitable context for functional
analysis, establishing the general existence of Feynman functors.
We also prove that any other Feynman functor is conjugated (in a very
nice way, which we call quasi essentially injective conjugation) to
that obtained extending the classic Feynman rules, establishing uniqueness.

There are two fundamental steps in showing that Feynman functors behave
as a bridge between reconstruction conjectures and superposition principles:
\begin{enumerate}
\item[s1)] \emph{showing that the reconstruction conjectures admit a sheaf-theoretic
characterization.} For each finite set $V$, let $\mathbf{S}_{\mathfrak{s},V}$
denote the category of hypergraphs which have structure of type $\mathfrak{s}$
and vertex set $V$. Varying $V$ we get a prestack $S_{\mathfrak{s}}$.
By making use of a deleting process, say $\mathcal{D}$, we get a
new prestack $\mathcal{D}S_{\mathfrak{s}}$ and a morphism $\mathcal{D}_{\mathfrak{s}}:S_{\mathfrak{s}}\Rightarrow\mathcal{D}S_{\mathfrak{s}}$,
i.e, a family of functors $\mathcal{D}_{\mathfrak{s},V}:\mathbf{S}_{\mathfrak{s},V}\rightarrow\mathcal{D}\mathbf{S}_{\mathfrak{s},V}$.
We show that different reconstruction conjectures consist in different
choices of $\mathcal{D}$ and that the corresponding $\mathcal{D}_{\mathfrak{s}}$
is objectwise essentially injective;
\item[s2)] \emph{proving that we can always consider Feynman functors which
are not only arbitrary functors, but actually monoidal and essentially
injective}. Monoidal property means that the analytic expression of
two disjoint structured hypergraphs is the product of the corresponding
analytic expressions. In turn, the essentially injectivity property
means that the hypergraphs are totally described by their analytic
expression.
\end{enumerate}
$\quad\;\,$Let $\mathbf{A}$ be a monoidal category endowed with
a structure of context for doing functional analysis, and let $Z_{V}:\mathbf{S}_{\mathfrak{s},V}\rightarrow\mathbf{A}_{\tau V}$
be a Feynman functor assigning to each hypergraph with $\mathfrak{s}$-structure
an analytic expression in $\mathbf{A}$. Let $\mathcal{D}$ be a deleting
process and suppose that we have a canonical morphism from $\mathcal{D}_{\mathfrak{s},V}G$
to the disjoint union of the parts of $G$ obtained via deleting (which
generally happens). Since $Z$ is monoidal, for each $G$ we have
a corresponding morphism 
\[
Z_{V}(G)\rightarrow\bigotimes_{\text{pieces}}Z(\text{pieces of }G).
\]

Suppose now that the reconstruction conjecture induced by $\mathcal{D}$
is satisfied. Then, because $Z_{V}$ is essentially injective, it
follows that two hypergraphs $G$ and $G'$ are isomorphic iff their
analytic expressions have the same decomposition in terms of the analytic
expressions of the pieces. This conclusion is precisely one example
of a superposition principle (in the sense axiomatized here). Let
us call it the $\mathcal{D}$\emph{-superposition principle. }Thus,
from s1) and s2) there follows our main result:
\begin{theorem}
\label{thm_introduction} The $\mathcal{D}$-reconstruction conjecture
for a prestack $S_{\mathfrak{s}}$ of $\mathfrak{s}$-structured hypergraphs
is true only if for any Feynman functor the $\mathcal{D}$-superposition
principle holds.
\end{theorem}
This theorem can be regarded both as an obstruction to the validity
of reconstruction conjectures and as a source of new superposition
principles. This relation becomes more involved when we think of the
role of quantum field theory. Indeed, consider $\mathbf{S}_{\mathfrak{s}}$
as the category of hypergraphs which parametrize the worldvolume of
particles, strings or branes (for instance, of Feynman graphs of QED
or some other gauge theory). Suppose we find $\mathcal{D}$ such that
the $\mathcal{D}$-reconstruction conjecture is true. Then each Feynman
functor (in particular that obtained by the Feynman rules of QED,
etc.) will produce a new superposition principle for the scattering
amplitudes.

On the other hand, since $\mathbf{S}_{\mathfrak{s}}$ are the Feynman
graphs of a physical theory, we can analyze whether these superposition
principles for the scattering amplitutes exist or not, looking at
concrete experiments of LHC, trying to find a counterexample. If found,
it will be a strong indicative that the $\mathcal{D}$-reconstruction
conjecture is false. Another approach is to notice that the existence
of new superposition principles in quantum theories produce many logic
implications \citep{superposicao_MQ_1,superposicao_MQ_2}, so that
assuming the validity of the $\mathcal{D}$-conjecture we could verify
if the induced logic implications contradict those that are experimentally
realized.

This paper is organized as follows. In Section \ref{sec_feynman}
we define what is meant by a $\mathfrak{s}$-structured hypergraph
and give many examples of objects that can be regarded as such. We
also show that the category of all categories that can be embedded
into $\mathbf{S}_{\mathfrak{s}}$ for some $\mathfrak{s}$ is complete.
In Section \ref{recontruction_conjectures} the sheaf-theoretic characterization
of the classical reconstruction conjecture is given and analogous
conjectures are defined, as needed for step s1). We also prove that
if one work with hypergraphs which contain labelings as part of their
structures, then many reconstruction conjectures are true. In Section
\ref{sec_functional_analysis} the notions of context for functional
analysis and analytic expressions are axiomatized and many examples
are given. In Section \ref{Feynman_functors} Feynman functors and
functorial Feynman rules are defined and the existence and uniqueness
up to quasi essentially injetive conjugation is established, as required
for step s2). We also show how to recover the classic Feynman rules
from this general approach. In Section \ref{sec_superposition} the
notion of superposition principle is formalized and a formal proof
of Theorem \ref{thm_introduction} is given. Finally, in Section \ref{applications}
some applications of our results on hypergraph theory, manifold topology
and perturbative quantum field theory are presented.
\begin{remark}
Along this article, by an oplax monoidal functor we mean one where
only the morphisms of the products are reverted. So, $F:\mathbf{C}\rightarrow\mathbf{C}'$
is oplax if it becomes endowed with natural transformations $F(X\otimes Y)\rightarrow F(X)\otimes'F(Y)$
and a morphism $1'\rightarrow F(1)$ making the appropriated diagrams
commutative.
\end{remark}

\section{Structured Hypergraphs\label{sec_feynman}}

There are several ways to define a hypergraph. For us, a \emph{hypergraph}
$G$ consists of a finite (possibly empty) set $V$ of vertices and
for each $j>1$ a finite (possibly empty) set $E_{j}$ of $j$-edges
and a $j$-adjacency function $\psi_{j}:E_{j}\rightarrow\operatorname{bin}(V,j)$
such that $\psi_{1}=id_{V}$, where $\operatorname{bin}(V,j)$ denotes
the set of $j$-subsets of $V$. We usually write $E_{1}=V$, so that
a $1$-edge is just a vertice. If $v\in\psi_{j}(e)$ we say that $e$
is adjacent to $e$. For each $v\in V$ and each $j>1$, let $d_{j}(v)$
denote the number of $j$-edges that are adjacent to $v$. Using these
notations, we have \citep{hypergraph_theory}: 
\begin{equation}
\sum_{j>1}\sum_{v}d_{j}(v)=\sum_{j>1}j\vert E_{j}\vert.\label{degrees_vertices_edges_formula}
\end{equation}

Let $\mathbb{N}$ be the set of natural number regarded as a discrete
category and notice that for each set $X$ the rule $j\mapsto\operatorname{bin}(X,j)$
extends to a functor $\operatorname{bin}(X,-):\mathbb{N}\rightarrow\mathbf{Set}$.
Thus, a hypergraph is equivalently a pair $(E,\psi)$, where $E:\mathbb{N}\rightarrow\mathbf{FinSet}$
and $\psi:E\Rightarrow\operatorname{bin}(E(1),-)$ is a natural transformation
such that $\psi_{1}=id$. Here, $\mathbf{FinSet}$ denotes the category
of finite sets. The equivalence between both definitions is obtained
via the identifications $E(j)=E_{j}$.

The rule assigning to each set $X$ its set $\operatorname{bin}(X,j)$
of $j$-subsets also extends to a functor $\operatorname{bin}(-,j):\mathbf{Set}\rightarrow\mathbf{Set}$.
We can then define a \emph{morphism} $f:G\rightarrow G'$ of between
hypergraphs $(E,\psi)$ and $(E',\psi')$ as a natural transformation
$f:E\Rightarrow E'$ which commutes with adjacencies, i.e, such that
$\psi'\circ f=\operatorname{bin}(f_{1},-)\circ\psi$. We have the
category $\mathbf{Hyp}$ of hypergraphs. Under the operation of taking
disjoint unions of hypergraphs it becomes a symmetric monoidal category
whose neutral object is the empty hypergraph.

Sometimes we will work with \emph{bounded hypergraphs}. We say that
$G$ in \emph{bounded} if there is some $b$ such that $E_{j}=\varnothing$
for each $j\geq b$. The smallest of these $b$'s is the \emph{bounding
degree }of $G$\emph{. }For each $b>0$ we have monoidal subcategories
$\mathbf{Hyp}^{b}\subset\mathbf{Hyp}$ of \emph{$b$-bounded hypergraphs}
with fixed bounding degree $b$. For instance, if $b=2$ this is the
category of what is known as finite pseudographs or finite graphs,
depending on the author and we will write $\mathbf{Grph}$ instead
of $\mathbf{Hyp}^{2}$.
\begin{remark}
Let $\mathbf{C}\subset\mathbf{Hyp}$ be some category of hypergraphs.
For fixed $V$ we can consider the full subcategory $\mathbf{C}_{V}\subset\mathbf{C}$
of hypergraphs in $\mathbf{C}$ whose vertex set is $V$ or empty.
Even if $\mathbf{C}$ is a monoidal subcategory, if $V\neq\varnothing$
then $\mathbf{C}_{V}$ is not monoidal. Indeed, if $G,G'$ have the
same vertex set $V$, then $G\sqcup G'$ has vertex set $V\sqcup V$.
This is one of the motivations for considering reconstruction conjectures
in a sheaf-theoretic perspective, as will be discussed in the next
section.
\end{remark}
$\quad\;\,$In the following we will work with \emph{structured hypergraphs},
in that for any $j\geq1$ we have functions $\varepsilon_{j}:E_{j}\rightarrow s_{j}$.
We think of $s_{j}$ as a set of $j$-structures in the set of $j$-edges.
Thus, $\varepsilon_{j}$ assigns to each $j$-edge a corresponding
$j$-structure. We form the category $\mathbf{S}\subset\mathbf{Hyp}$
whose morphisms are hypergraph morphisms $f:G\rightarrow G'$ preserving
$j$-structures. In more precisely terms, for each functor $s:\mathbb{N}\rightarrow\mathbf{Set}$,
called a \emph{functor of structures}, we define a category $\mathbf{S}_{s}$
as follows. Objects are \emph{$s$-structured hygraphs}, i.e, pairs
$(G,\varepsilon)$, where $G=(E,\psi)$ is a hypergraph and $\varepsilon:E\Rightarrow s$
is a natural transformation. Morphisms $f:(G,\varepsilon)\rightarrow(G',\varepsilon')$
are hypergraph morphisms between the underlying hypergraphs such that
$\varepsilon'\circ\operatorname{bin}(f,-)=\varepsilon$. If we are
working with bounded hypergraphs we can consider $\mathbf{S}^{b}\subset\mathbf{Hyp}^{b}$,
defined analogously.

Given $s$-structured hypergraphs $(G,\varepsilon)$ and $(G',\varepsilon')$,
the disjoint union $G\sqcup G'$ can be naturally regarded as a $s$-structured
hypergraph with the transformation $(\varepsilon\sqcup\varepsilon')_{j}:E_{j}\sqcup E'_{j}\rightarrow s_{j}$
given by the composition below, where the second map is the codiagonal.
Also, the empty hypergraph has a unique $s$-structure, so that $\mathbf{S}\subset\mathbf{Hyp}$
is actually a monoidal subcategory.$$
\xymatrix{\ar@/_{0.3cm}/[rr]_{(\varepsilon\sqcup\varepsilon')_{j}} E_{j}\sqcup E'_{j} \ar[r]^-{\varepsilon _j \sqcup \varepsilon ' _j} & s_{j} \sqcup s_{j} \ar[r] & s_{j}}
$$
\begin{remark}
\label{remark_additive_structures} When $s_{j}:\mathbb{N}\rightarrow\mathbf{Ab}$,
i.e, when the structural functor takes values in the category of abelian
groups, the map $(\varepsilon\sqcup\varepsilon')_{j}$ can be identified
with $\varepsilon_{j}+\varepsilon'_{j}$. In these situations we say
that $s$ is an \emph{additive structure}.\emph{ }In this paper, essentially
all structures will take values in $\mathbf{FinSet}$, so that we
can always think of them as additive structures by replacing a finite
set $[n]$ with the abelian group $\mathbb{Z}_{n}$ that it generates.
\end{remark}
The last construction can be easily generalized by considering not
only one functor of structure $s$, but a family of them. In fact,
let $\mathfrak{s}:\mathbb{N}\rightarrow[\mathbb{N};\mathbf{Set}]$
be a functor that to each $k$ assigns a structure functor $\mathfrak{s}_{k}:\mathbb{N}\rightarrow\mathbf{Set}$,
so that $\mathfrak{s}_{k,j}\in\mathbf{Set}$ (equivalently, $\mathfrak{s}$
can be regarded as a bifunctor $\mathbb{N}\times\mathbb{N}\rightarrow\mathbf{Set}$)\footnote{Here $[\mathbf{C};\mathbf{D}]$ denotes the functor category.}.
Let $\operatorname{cst}_{E}:\mathbb{N}\rightarrow[\mathbb{N};\mathbf{Set}]$
be the constant functor in $E$. We define a \emph{$\mathfrak{s}$-structured
hypergraph} as previously: it is a pair $(G,\epsilon)$, but now $\epsilon$
is a natural transformation $\epsilon:\operatorname{cst}_{E}\Rightarrow\mathfrak{s}$.
Thus, it is a rule that to each $k$ assigns a natural transformation
$\epsilon_{k}:E\Rightarrow s_{k}$, which means that for fixed $j\geq1$
we have a family $\epsilon_{k,j}:E_{j}\rightarrow\mathfrak{s}_{k,j}$.
Morphisms of $\mathfrak{s}$-structured hypergraphs also are defined
analogously, so that we have a monoidal category $\mathbf{S}_{\mathfrak{s}}$.
If we are working with bounded hypergraphs we have the corresponding
monoidal category $\mathbf{S}_{\mathfrak{s}}^{b}$.

\subsection{Embedded Subcategories}

In this subsection we will discuss some category of hypergraphs that
can be embedded into the category of structured hypergraphs in an
essentially injective way. More precisely, we will give examples of
subcategories $\mathbf{C}\subset\mathbf{Hyp}$ such that for each
fixed $V$ the corresponding $\mathbf{C}_{V}$ can be realized, up
to equivalence, as a full subcategory of $\mathbf{S}_{\mathfrak{s},V}$
for some $\mathfrak{s}$, meaning that there exists a fully faithful
(and, therefore, essentially injective) inclusion functor $\imath:\mathbf{C}_{V}\hookrightarrow\mathbf{S}_{\mathfrak{s},V}$.
\begin{example}[colouring]
 Recall that a (vertex) \emph{colouring} for a hypergraph $G$ consists
of a finite set $A\subset\mathbb{N}$ of colors and a function $c:V\rightarrow S$
(assigning to each vertice its color) such that\emph{ }each hyperedge
contains at least two vertices of distinct colors. In other words,
for each $j$ the composition $c_{j}=\operatorname{bin}(V,c)\circ\psi_{j}$
is non-constant. We will work with $A=[n]=\{1,2,...,n\}$. Notice
that for graphs we recover the usual notion of graph coloring. A \emph{$n$-colored
}hypergraph is one in which a coloring with a set of $n$ colors was
fixed, i.e, it is a pair $(G,c)$ with $c:V\rightarrow[n]$. A morphism
$f:(G,c)\rightarrow(G',c')$ of $n$-colored hypergraphs is a hypergraph
morphism which preserves the coloring, i.e, $c'\circ f=c$. Denote
by $n\mathbf{Col}$ the category of $n$-colored hypergraphs. Define
$\mathfrak{s}$ by $\mathfrak{s}_{1,1}=[n]$ and $\mathfrak{s}_{j,k}=0$
if $j\neq1$ or $k\neq1$. Let $n\mathbf{S}_{\mathfrak{s},V}$ be
the full subcategory of $\mathbf{S}_{\mathfrak{s},V}$ whose objects
are $\mathfrak{s}$-structured graphs $(G,\epsilon)$ with $\epsilon_{1,1}=c$
and $\epsilon_{j,k}=0$ if $j\neq1$ or $k\neq1$. The condition $\epsilon'\circ f=\epsilon$
means precisely that $c'\circ f=c$, so that $n\mathbf{Col}_{V}\simeq\mathbf{S}_{\mathfrak{s},V}$.
\begin{example}[labeled hypergraphs]
 \label{ex_labeled_hypergraph} By a \emph{labeling} of a hypergraph
$G$ we mean a bijection $\varphi_{1}:V\rightarrow[\vert V\vert]$,
where $[n]=\{1,...,n\}$. Notice that this map induces bijections
$\operatorname{bin}(\varphi_{1},j)$ for every $j>1$ and, by restriction,
corresponding bijections $\varphi_{j}:E_{j}\rightarrow[\vert E_{j}\vert]$.
A \emph{labeled hypergraph} is one in which a labeling has been chosen,
i.e, it is a pair $(G,\varphi_{1})$. A morphism of labeled hypergraphs
$(G,\varphi_{1})$ and $(G',\varphi'_{1})$ is a triple $(f,\alpha,\alpha')$,
where $f:G\rightarrow G'$ is a morphism of hypergraphs and $\alpha$
and $\alpha'$ are families of maps such that the diagram below commutes.
Notice that, since $\varphi_{j}$ and $\varphi_{1}$ are bijective,
$\alpha$ and $\alpha'$ are actually determined by $f$, $\varphi$
and $\varphi'$.$$
\xymatrix{\ar[d]_{\alpha _j} [\vert E_j] \vert \ar[r]^{\varphi _j} & \ar[d]_{\psi _j} E_j \ar[r]^{f_j} & \ar[d]^{\psi' _j} E'_j & [\vert E'_j] \vert \ar[l]_{\varphi ' _j} \ar[d]^{\alpha '_j} \\
\operatorname{bin}([\vert V \vert ],j)) \ar[r]_{\operatorname{bin}(\varphi _1,j)} & \ar[r]_{\operatorname{bin}(f_1,j)} \operatorname{bin}(V,j)) & \operatorname{bin}(V',j)) & \operatorname{bin}([\vert V' \vert ],j)) \ar[l]^{\operatorname{bin}(\varphi _1 ',j)} }
$$ We then have a category $\mathbf{Hyp^{\ell}}$ of labeled hypergraphs.
Let us show that for every fixed finite set $V$ we have a canonical
functor $\imath:\mathbf{Hyp}_{V}^{\ell}\hookrightarrow\mathbf{S}_{\mathfrak{s},V}$
for some $\mathfrak{s}$. From the above will then follows that $\imath$
is essentially injective, as desired. We define $\mathfrak{s}_{k,j}$
as the group $\mathbb{Z}_{\vert V\vert}$, if $j=k=1$, and the trivial
group otherwise, i.e, if $j\neq1$ or $k\neq1$. Now, for every labeled
hypergraph $(G,\varphi)$ with vertex set $E(1)=V$, take $\epsilon_{k,j}:E_{j}\rightarrow\mathfrak{s}_{k,j}$
as the trivial map if $k\neq1$ or $j\neq1$, and $\epsilon_{1,1}=\varphi_{1}$.
For a morphism $f$ of labeled graphs, it is clear that $\epsilon'_{k,j}\circ\operatorname{bin}(f,j)=\epsilon_{k,j}$,
so that it becomes a morphism of structured graphs, and thus $\mathbf{Hyp}_{V}^{\ell}\hookrightarrow\mathbf{S}_{\mathfrak{s},V}$.
\end{example}
\end{example}
\begin{remark}
\label{isomorphism_labeled_graphs} If $(G,\varphi)$ and $(G',\varphi')$
are labeled hypergraphs such that $\vert E_{j}\vert=\vert E'_{j}\vert$
for some $j$, then it follows from the commutativity of the last
diagram that any morphism $f:G\rightarrow G'$ has the $j$th component
$f_{j}$ completely determined by the labelings, i.e, $f_{j}=\varphi_{j}\circ\varphi_{j}^{-1}$.
In particular, it is a bijection. Consequently, if there exists an
isomorphism between two labeled graphs, then it is unique. Furthermore,
if we are working in $\mathbf{Hyp}_{V}^{\ell}$, then we can always
assume $f_{1}=id_{V}$.
\end{remark}
\begin{example}[Feynman graphs]
 \label{ex_feynman_graph}A \emph{Feynman graph} is a graph $G$
that has further structure satisfying additional properties. Precisely,
it has a decomposition $V=V^{0}\sqcup V^{1}$ into \emph{external
}(or \emph{false})\emph{ vertices} and \emph{internal }(or \emph{fundamental})\emph{
vertices}, respectively, and a function $g:V\rightarrow\mathbb{Z}_{\geq0}$,
called \emph{genus map}. This data is required to satisfy:
\begin{enumerate}
\item[c1)] if $v\in V^{0}$, then $d(v)=1$ and $g(v)=0$;
\item[c2)] there are no edges between two external vertices.
\end{enumerate}
From the conditions above we see that there exists $E^{0}\subset E$
with $\vert E^{0}\vert=\vert V^{0}\vert$, so that we also have a
decomposition $E=E^{0}\sqcup E^{1}$ into \emph{external edges} (or
\emph{tails}) and \emph{internal edges}, respectively, where $E^{1}=E-E^{0}$.
A \emph{morphism} of Feynman graphs is a hypergraph morphism preserving
the additonal structure, i.e, such that $f_{1}$ maps $V^{i}$ into
$V'^{i}$ and $g'\circ f_{1}=g$. Consequently, $f_{2}$ preserves
$E^{0}$ and, therefore, $E^{1}$. Let $\mathbf{Fyn}$ denote the
category of Feynman graphs. Notice that to give a decomposition of
$V$ is equivalent to giving a surjective function $\pi:V\rightarrow\mathbb{Z}_{2}$
by $\ensuremath{\pi_{V}^{-1}(i)=V^{i}}$. Take 
\[
\mathfrak{s}_{k,1}=\begin{cases}
\mathbb{Z}_{2}, & k=1\\
\mathbb{Z}_{\geq0}, & k=2\\
0, & k>2
\end{cases}
\]
with $\mathfrak{s}_{k,j}=0$ if $j>1$. If $(G,\pi,g)$ is a Feynman
graph on $V$, define $\epsilon_{1,1}=\pi$, $\epsilon_{2,1}=g$ and
$\epsilon_{k,j}=0$, otherwise. For a given graph morphism $f$ the
condition $\epsilon'_{k,j}\circ\operatorname{bin}(f,j)=\epsilon_{k,j}$
means precisely that $f$ preserves $\pi$ and $g$. Thus, $f$ is
a Feynman graph morphism iff it is a morphism of $\mathfrak{s}$-structured
graphs. This gives the desired embedding $\mathbf{Fyn}_{V}\hookrightarrow\mathbf{S}_{\mathfrak{s},V}.$
\end{example}
\begin{remark}
If to conditions c1) and c2) above we add
\begin{enumerate}
\item[c3)] if $g(v)=0$, then $d(v)\geq3$;
\item[c4)] if $g(v)=1$, then $d(v)\geq1$,
\end{enumerate}
then we have what is known as \emph{stable graphs}, since this class
of graphs contains those arising in the study of stable curves \citep{Deline-Munford_1,Costello}.
If $\mathbf{Stb}\subset\mathbf{Fyn}$ is the full subcategory of stable
graphs, then $\mathbf{Stb}_{V}$ can also be embedded in $\mathbf{S}_{\mathfrak{s},V}$
for the same $\mathfrak{s}$ as $\mathbf{Fyn}_{V}$.

\end{remark}
\begin{example}[structured Feynman hypergraphs]
 \label{ex_struct_Feynman}Feynman graphs can be generalized in two
directions: allowing hyperedges and allowing additional structures.
In the first case we say that we have a \emph{Feynman hypergraph},
while in the second one we say that we have a \emph{structured Feynman
graph}. It is straightforward to verify (following the same kind of
construction used in the previous examples) that these classes of
graphs produce categories $\mathbf{HypFyn}_{V}$ and $\mathbf{Fyn}_{\mathfrak{s},V}$
which can be embedded in $\mathbf{S}_{\mathfrak{s},V}$ for some $\mathfrak{s}$.
Special examples are the so-called \emph{generalized Feynman graphs
}\citep{generalized_Feynman_1,generalized_Feynman_2-2} and \emph{sectored
Feynman graphs} \citep{sectored_feynman_1,sectored_Feynman_2,sectored_feynman_3}\footnote{The terminology is not standard: some authors use generalized Feynman
graphs to refer to sectored Feynman graphs. There is also a notion
of \emph{generalized Feynman amplitudes}, introduced in \citep{generalized_Feynman_3},
which (to the best of the author's knowledge) it is not directly related
with the other ones.}. For instance, in generalized Feynman graphs we have an additional
decomposition of $V^{1}$ into \emph{positive} and \emph{negative}
vertices, i.e, $V^{1,+}\sqcup V^{1,-}$ (which means that now we need
to use a projection $\pi:V\rightarrow\mathbb{Z}_{3}$ onto $\mathbb{Z}_{3}$
instead of onto $\mathbb{Z}_{2}$) and the edge set $E$ is constrained
by the condition that there is no edges between vertices of the same
signal.
\begin{example}[ribbon graphs]
 An alternative way of defining a graph $G$ is as being given by
a set $V$ of vertices, a set $E$ of edges, an incidence function
$s:V\rightarrow E$ and an involution $i:E\rightarrow E$ without
fixed points. Morphisms $f:G\rightarrow G'$ are defined as functions
$f_{V}:V\rightarrow V'$ and $f_{E}:E\rightarrow E'$ such that $f_{E}\circ s=s'\circ f_{V}$
and $f_{E}\circ i=i'\circ f_{E}$. Denote by $\mathbf{Grph'}$ this
category. So, $\mathbf{Grph'}\simeq\mathbf{Hyp}^{2}$. Of special
interest is the subcategory $\mathbf{Ribb}$ of \emph{ribbon} (or
\emph{fat}) graphs. They become endowed with a permutation $\sigma:E\rightarrow E$
satisfying the following property:
\end{example}
\begin{enumerate}
\item[r)] let $\langle\sigma\rangle$ be the cyclic group generated by $\sigma$.
It acts on $E$ giving a decomposition, which must coincide with that
induced by the fibers $s^{-1}(x)$ of $s$.
\end{enumerate}
The morphisms between ribbon graphs are graph morphisms which preserve
the permutations. Let $\mathbf{Ribb}_{0}$ be the category of graphs
with a permutation $\sigma:E\rightarrow E$ but which does not necessarily
satisfies r). It is clear that $\mathbf{Ribb}_{0}\hookrightarrow\mathbf{S}_{\mathfrak{s}}^{1}$
for some $\mathfrak{s}$. But $\mathbf{Ribb}$ is a full subcategory
of $\mathbf{Ribb}_{0}$, so that it can also be embedded into a category
of structured graphs.
\end{example}
The intersection of two embedded subcategories of structured hypergraphs
remains a category of structured hypergraphs. More precisely, if $\mathbf{C}\hookrightarrow\mathbf{S}_{\mathfrak{s}}$
and $\mathbf{C}'\hookrightarrow\mathbf{S}_{\mathfrak{s}'}$, then
$\mathbf{C}\cap\mathbf{C}'$ can be embedded in both $\mathbf{S}_{\mathfrak{s}}$
and $\mathbf{S}_{\mathfrak{s}'}$. For instance, in the last example
$\mathbf{Fyn}_{\mathfrak{s}}=\mathbf{Fyn}\cap\mathbf{S}_{\mathfrak{s}}$.
More generally, arbitrary limits of categories of structured hypergraphs
remain a category of structured hypergraphs. In fact, let $\mathfrak{S}$
be the category defined as follows. Its objects are categories $\mathbf{C}$
such that there exists a functor of structures $\mathfrak{s}$ and
an essentially injective embedding $\imath:\mathbf{C}\hookrightarrow\mathbf{S}_{\mathfrak{s}}$,
while the morphisms are functors. So $\mathbf{\mathfrak{S}}$ it is
actually a full subcategory of $\mathbf{Cat}$. Since full embeddings
are monadic, it follows that they reflect limits \citep{handbook_category}.
Thus:
\begin{proposition}
\label{limits_structured_category}The category $\mathfrak{S}$ is
complete.
\end{proposition}
We end this section with a convention which will be specially important
in the construction of Feynman functors in Section \ref{Feynman_functors}.
\begin{remark}
\label{remark_convention_labeling} Let $(G,\varphi)$ and $(G',\varphi')$
two labeled hypergraphs with vertex set $V$. Regarding them as structured
hypergraphs as done in Example \ref{ex_labeled_hypergraph} we find
ambiguities when considering the induced labeling in $G\sqcup G'$.
In order to fix this we will use the following convention: let $[\vert V\vert]'=\{1',...,\vert V\vert'\}$
be a copy of $[\vert V\vert]$ and in the disjoint union
\[
[\vert V\vert]\sqcup[\vert V\vert]'=\{1,...,\vert V\vert,1',...,\vert V\vert'\}
\]
consider the ordering 
\[
1\leq2\leq...\leq\vert V\vert\leq1'\leq2'\leq...\leq\vert V\vert'.
\]
Define in $V\sqcup V$ a similar ordering, i.e, $\varphi(1)\leq...\leq\varphi(\vert V\vert)\leq\varphi(1')\leq...\leq\varphi(\vert V\vert').$
Then define the labeling in $G\sqcup G'$ as the unique bijection
such that this ordering is preserved.
\end{remark}

\section{Reconstruction Conjectures\label{recontruction_conjectures}}

Given two sets $V,V'$, with $V'\subset V$, let $V'^{c}=V-V'$ denote
the complement of $V'$ in $V$. We have a pair of adjoint functors
\[
D_{V',V}:\mathbf{Hyp}_{V}\rightleftharpoons\mathbf{Hyp}_{V'}:I_{V',V},
\]
defined as follows. The right adjoint $I_{V',V}$ is just the inclusion
functor. More precisely, if $G$ is a hypergraph with vertex set $V'$,
then $I_{V',V}(G)$ is the graph obtained by adding the elements of
$V'^{c}$ as isolated vertices. On the other hand, $D_{V',V}$ is
the functor that takes a hypergraph $G$, with vertex set $V$, and
delete the vertices $V'^{c}$ together with their adjacent hyperedges.

So, if we fix a set $X$ and define $\mathbb{X}$ as the category
whose objects are subsets $V\subset X$ and whose morphisms are inclusions,
varying $V,V'$ we get functors $I:\mathbb{X}\rightarrow\mathbf{Cat}$
and $D:\mathbb{X}^{op}\rightarrow\mathbf{Cat}$. Notice that the process
of adding (resp. deleting) a finite number of vertices is equivalent
to iterating the process of adding (resp. deleting) a single vertice.
This means that we have a distinguished class $J\subset\mathrm{Mor}(\mathbb{X})$
of morphisms, given by inclusions $\jmath_{x}:V'\hookrightarrow V$,
with $V'=V-x$ for some $x\in V$. We can think of $J$ as a rule
assigning to each $V\in\mathbb{X}$ the collection of morphisms $J(V)=(\jmath_{x})_{x\in V}$,
which we call the \emph{covering family} \emph{of $V$}. It is clear
that if $V'\subset V$, then $V'-x\subset V-x$. Since the only morphisms
in $\mathbb{X}$ are inclusions and the only covering families are
$(\jmath_{x})$, the previous condition implies that given a covering
family $J(V)$ and a morphism $V'\rightarrow V$, then for any covering
family $J(V')$ and each $\jmath'_{x'}\in J(V')$ there exists some
$\jmath_{x}\in J(V)$ (actually $\jmath_{x'}$) such that the diagram
below commutes.$$
\xymatrix{\ar[d]_{\jmath _{x'}'} V'-x' \ar[r]^{f-x'} & V-x' \ar[d]^{\jmath _x'} \\
V' \ar[r]_f & V }
$$ 

In other words, $J$ is a coverage for $\mathbb{X}$. Therefore, we
can talk about stacks on the site $(\mathbb{X},J)$. More precisely,
we have a reflexive subcategory of presheaves
\[
\imath:\mathbf{Stack}(\mathbb{X},J)\rightleftharpoons[\mathbb{X}^{op};\mathbf{Cat}]:L
\]
whose reflection $L$ preserve finite limits. This reflexive subcategory
is just the localization of the presheaf category at the local isomorphism
system associated with $J$. If $\jmath\in J$, then $Y(\jmath)$
belongs to this system, where $Y$ denotes the Yoneda embedding. Therefore,
$LY(\jmath)$ is an isomorphism, so that from Yoneda lemma $LF(\jmath)$
is an isomorphism for every $F:\mathbb{X}^{op}\rightarrow\mathbf{Cat}$.
Particularly, it is for the deleting functor $D$ above and for every
subfunctor $C\subset D$. This means that \emph{after localization,
the process of} \emph{deleting vertices becomes an equivalence}. Furthermore:
the same holds for every subfunctor $C\subset D$. Such a subfunctor
assigns to each $V\subset X$ a subcategory $\mathbf{C}_{V}\subset\mathbf{Hyp}_{V}$
which is invariant by vertex deleting, i.e, we get an induced functor
$C_{V',V}:\mathbf{C}_{V}\rightarrow\mathbf{C}_{V-V'}$. So, after
localizing, deleting vertices is an equivalence independently of the
class of hypergraphs considered.

We should not expect the same result before localizing, since in general
there are much more graphs with $\vert V\vert$ vertices than graphs
with $\vert V\vert-1$ vertices. But, we can ask if in a given class
of graphs, i.e, for a given subfunctor $C\subset D$, for every $\jmath_{x}\in J$
the corresponding functor $C(\jmath_{x}):\mathbf{C}_{V}\rightarrow\mathbf{C}_{V-x}$
is at least essentially injective. This remains a very strong requirement,
since we are asking if any information of a hypergraph $G\in\mathbf{C}_{V}$
can be recovered from the information after deleting a single \emph{fixed}
vertice $x\in V$. Thus, we can think of taking all $x\in V$ into
account simultaneously. More precisely, we can ask if the induced
composition below is essentially injective.$$
\xymatrix{ \ar@/^{0.5cm}/[rrr]^{\Delta C_{V}} \mathbf{C}_V \ar[r]_-{\Delta} & \prod _{x\in V} \mathbf{C}_V \ar[rr]_-{\prod _{x\in V}C(\jmath _x)} && \prod _{x\in V} \mathbf{C}_{V-x}}
$$

This is just a sheaf theoretically formulation of what is usually
known in hypergraph theory as the Reconstruction Conjecture for the
class of graphs defined by $C$. In fact, calling an isomorphism in
the image of $\Delta C_{V}$ a \emph{hypomorphism}, the assertion
that $\Delta C_{V}$ is essentially injective is equivalent to:
\begin{conjecture}[RC-$C$]
 \emph{Two hypergraphs in $\mathbf{C}_{V}$ are isomorphic iff they
are hypomorphic.}
\end{conjecture}
\begin{remark}
Of course, if the categories $\mathbf{C}_{V}$ actually belong to
$\mathbf{Grph}_{V}$, i.e, if we are in the context of graphs instead
of general hypergraphs, then the above discussion reproduces the same
conjecture, but now in graph theory. Some consequences of this categorical
description (not directly related with the sheaf structure and specially
concerning obstructions to the existence of non-nilpotent graph invariants)
are in a work in preparation.
\end{remark}

\subsection{Disjoint Reconstruction \label{sec_other_conjectures}}

In the last section we gave a sheaf-theoretically description of the
classical graph reconstruction conjecture. Our approach, on the other
hand, has a problem:
\begin{itemize}
\item the prestack $D$ is morphismwise adjoint to $I$, so that it is natural
to believe that $I$ should appear in any fundamental construction
involving $D$. However, it was not used in the construction of $\Delta D_{V}$.
\end{itemize}
$\quad\;\,$In order to fix these pathologies, notice that despite
of $I:\mathbb{X}\rightarrow\mathbf{Cat}$ not being strong monoidal
(since for arbitrary $V,V'$ we do not have an equivalence between
$\mathbf{Hyp}_{V\sqcup V'}$ and $\mathbf{Hyp}_{V}\times\mathbf{Hyp}_{V'}$),
it is lax comonoidal (because for generic $V,V'$ there are more hypergraphs
over $V\sqcup V'$ than pairs of hypergraphs over $V$ and $V'$).
Consequently, we have a natural transformation 
\[
\xi_{V',V}^{I}:\mathbf{Hyp}_{V}\times\mathbf{Hyp}_{V'}\rightarrow\mathbf{Hyp}_{V\sqcup V'}
\]
sending pairs $(G,G')$ into its disjoint union $G\sqcup G'$. We
introduced the upper index to emphasize that $\xi^{I}$ depends on
$I$. So, instead of $\Delta D_{V}$ we can consider the composition
$dD_{V}$ below.\begin{equation}{\label{dRC_diagram}
\xymatrix{ \ar@/_{0.5cm}/[rrrd]_{dD_V} \ar@/^{0.5cm}/[rrr]^{\Delta D_{V}} \mathbf{Hyp}_{V} \ar[r]_-{\Delta} & \prod _{x\in V} \mathbf{Hyp}_{V} \ar[rr]_-{\prod _{x\in V}D(\jmath _x)} && \prod _{x\in V} \mathbf{Hyp}_{V-x} \ar[d]^{\xi ^I} \\
&&& \mathbf{Hyp}_{\coprod _{x}(V-x)}}}
\end{equation}

This clearly fixes the initial problem for $D$, but this does not
makes sense for arbitrary $C\subset D$, since $C_{V,V'}$ may not
commute with $\xi_{V',V}^{I}$. That is, given $G\in\mathbf{C}_{V}$
and $G'\in\mathbf{C}_{V'}$ there is no guarantee that $G\sqcup G'\in\mathbf{C}_{V\sqcup V'}$.
If $C\subset D$ is a subfunctor satisfying this condition we will
say that it is \emph{proper}. In this case, for every $V$ the map
$dC_{V}$ is well defined.
\begin{example}[structured presheaves]
 \label{structured_presheaf_proper}A prestack of $\mathfrak{s}$-structured
hypergraphs is proper. More precisely, given any functor $\mathfrak{s}:\mathbb{N}\times\mathbb{N}\rightarrow\mathbf{Set}$,
the preshaf $S_{\mathfrak{s}}$ that to any $V$ assigns $\mathbf{S}_{\mathfrak{s},V}$
is proper. Indeed, given $(G,\epsilon)$ and $(G',\epsilon')$ in
$\mathbf{S}_{\mathfrak{s},V}$ and $\mathbf{S}_{\mathfrak{s},V'}$,
respectively, we can always introduce structure $\epsilon\sqcup\epsilon'$
in $G\sqcup G'$. In particular, the presheaf $S^{\ell}$ such that
$S^{\ell}(X)=\mathbf{Hyp}_{X}^{\ell}$ is proper.
\begin{example}
The trivial prestack $\varnothing$ such that $\varnothing_{V}=\varnothing$
for every $V$ is trivally proper.
\end{example}
\end{example}
\begin{remark}
Beware that subfunctors of proper subfunctors need not be proper.
That is, if $C\subset D$ is proper, then for arbitrary $C'\subset C$
it is not true that $C'\subset D$ is proper. On the other hand, the
intersection $S_{\mathfrak{s}}^{\ell}=S^{\ell}\cap S_{\mathfrak{s}}$
is proper, because $S_{\mathfrak{s}}^{\ell}=S_{\mathfrak{s}'}$ for
some $\mathfrak{s}'$.
\end{remark}
$\quad\;\,$Knowing that the problem is fixed by replacing $\Delta C$
with $\mathbf{d}C$, we can think of a conjecture somewhat analogous
to $C$-RC. In order to do this, let us say that the isomorphisms
in the image of each $\mathbf{d}C_{V}$ are \emph{weak hypomorphisms}.
So, we can then conjecture the Disjoint Reconstruction Conjecture:
\begin{conjecture}[dRC-$C$]
\emph{Two hypergraphs in a proper presheaf $C\subset D$ are isomorphic
iff they are weakly hypomorphic. In other words, $\mathbf{d}C_{V}$
is essentially injective.}
\end{conjecture}
\begin{remark}
From now on, if $C$ is any prestack of hypergraphs, we will use the
following simplified notations:
\begin{enumerate}
\item $\sqcap\mathbf{C}_{V}$ instead of $\prod_{x\in V}\mathbf{C}_{V-x}$;
\item $\sqcup\mathbf{C}_{V}$ instead of $\coprod_{x\in V}\mathbf{C}_{V-x}$;
\item $\mathbf{C}_{dV}$ instead of $\mathbf{C}_{\coprod_{x\in V}V-x}$.
\end{enumerate}
\end{remark}

\subsection{Category of Reconstruction Conjectures}

Let us now see that both conjectures RC and dRC can be considered
in an axiomatic background. For doing this we define a \emph{hypergraph
reconstruction context} as\emph{ }given by the following data:
\begin{enumerate}
\item for each set $X$ we have a subcategory $\mathfrak{C}\subset[\mathbb{X}^{op};\mathbf{Cat}]$
of \emph{applicable prestacks}, such that if $C\in\mathfrak{C}$,
then $C_{V}\subset\mathbf{Hyp}$ for each $V\subset X$;
\item a functor $\mathcal{D}:\mathfrak{C}\rightarrow[\mathbb{X}^{op};\mathbf{Cat}]$,
playing the role of a ``deleting process'' and that to any applicable
prestack $C$ it assigns the \emph{prestack of pieces $\mathcal{D}C$},
such that for each $V\subset X$ we have the \emph{category of pieces
}$\mathcal{D}C_{V}$\emph{;}
\item a natural transformation $\gamma:\imath_{\mathfrak{C}}\Rightarrow\mathcal{D}$,
where $\imath_{\mathfrak{C}}$ is the inclusion of applicable prestacks
into the category of prestacks.
\end{enumerate}
We will represent the reconstruction context simply by its deleting
process $\mathcal{D}$, except when we need more details. We say that
two hypergraphs are $\mathcal{D}$-\emph{hypomorphic }if their image
by $\mathcal{D}C_{V}$ are isomorphic. So, given a reconstruction
context $\mathcal{D}$ and an applicable prestack $C\in\mathfrak{C}$
we can consider the following $\mathcal{D}$-reconstruction conjecture
for $C$.
\begin{conjecture}[$\mathcal{D}$-RC-$C$]
\emph{ For each $V$, two hypergraphs in $C_{V}$ are isomorphic
iff they are $\mathcal{D}$-hypomorphic, i.e, each functor $\gamma_{V}:C_{V}\rightarrow DC_{V}$
is essentially injective.}
\end{conjecture}
Before giving examples, two important remarks:
\begin{remark}
\label{concrete_prestack} In the previous sections, the prestacks
of hypergraphs $C:\mathbb{X}^{op}\rightarrow\mathbf{Cat}$ were such
that for each $V\subset X$ the corresponding $C_{V}$ is not only
an arbitrary subcategory of $\mathbf{Hyp}$, but actually a subcategory
of $\mathbf{Hyp}_{V}$, i.e, we worked with prestacks that assign
to each $V$ a category of hypergraphs with vertex set $V$. From
now on, we will work with prestacks which a priori take values only
in $\mathbf{Hyp}$. This will be specially important in proving existence
and uniqueness of Feynman rules. In order to distinguish between these
situations, we will say that $C$ is a \emph{concrete prestack} if
$C_{V}\subset\mathbf{Hyp}_{V}$, using the bold notation $\mathbf{C}_{V}$
instead of $C_{V}$ (as we have used in previous sections).
\begin{remark}
\label{proper_prestack} We say that an applicable prestack $C\in\mathfrak{C}$
is \emph{proper} if it becomes endowed with a natural transformation
$\xi:C_{V}\times C_{V'}\rightarrow C_{V\sqcup V'}$. Recall that in
the last sections we defined a proper (concrete) prestack as such
that $\xi:(G,G')\mapsto G\sqcup G'$ is well defined. This means that
any proper concrete prestack (in the older context) is proper (in
the newer sense). However, the reciprocal is not true, due to the
last remark. In order to emphasize that this new concept is more general,
we will call them \emph{concretely proper} prestacks.
\end{remark}
\end{remark}
\begin{example}[RC and dRC]
 \label{RC_dRC_contexts}In order to recover the classical reconstruction
conjecture, take $\mathfrak{C}$ as the whole category of prestacks,
$\mathcal{D}C_{V}=\sqcap\mathbf{C}_{V}$ and $\gamma_{V}=\Delta\mathbf{C}_{V}$.
For recovering the disjoint reconstruction conjecture, take $\mathfrak{C}$
as the subcategory of proper prestacks, $DC_{V}=\mathbf{C}_{dV}$
and $\gamma_{V}=d\mathbf{C}_{V}$.
\begin{example}[trivial context]
 We also have a trivial reconstruction context $\mathcal{I}$, in
which $\mathfrak{C}$ is the whole category of prestacks and $\mathcal{D}$
and $\gamma$ are the identity functors. Notice that in it any recontruction
conjecture is satisfied.
\begin{example}[restrictions]
 Let $\mathcal{D}$ be a reconstruction context with category of
applicable prestacks $\mathfrak{C}$ and natural transformation $\gamma:\imath_{\mathfrak{C}}\Rightarrow\mathcal{D}$.
For any subcategory $\mathfrak{D}\subset\mathfrak{C}$ we get a context
$\mathcal{D}\vert_{\mathfrak{D}}$ by restricting $\mathcal{D}$ and
$\gamma$ to $\mathfrak{D}$.
\end{example}
\end{example}
\end{example}
We will build some categories of reconstruction conjectures, allowing
us to compare two of different conjectures. A \emph{left morphism
}between two reconstruction contexts $\mathcal{D}$ and $\mathcal{D}'$
is given by a functor $F:\mathfrak{C}\rightarrow\mathfrak{C}'$ between
the categories of applicable prestacks, together with a natural transformation
$\xi:\mathcal{D}\Rightarrow\mathcal{D}'\circ F$ between the deleting
process, such that the first diagram below commutes, i.e, if for every
$C\in\mathfrak{C}$ we have $\xi_{C}\circ\gamma_{C}=\gamma'{}_{F(C)}$.
This gives us a category $L\mathbf{RC}$. By inverting the direction
of $\xi$ we define \emph{right morphisms}, which produce a dual category
$R\mathbf{RC}$, characterized by the second diagram below.\begin{equation}{\label{morphism_conjectures}
\xymatrix{ \ar@{=>}[d]_{\imath _{F}} \imath _{\mathfrak{C}} \ar@{=>}[r]^{\gamma} & \mathcal{D} \ar@{=>}[d]^{\xi} &  \ar@{=>}[d]_{\imath _{F}} \imath _{\mathfrak{C}} \ar@{=>}[r]^{\gamma} & \mathcal{D} \\
\imath _{\mathfrak{C'}\circ F} \ar@{=>}[r]_{\gamma ' \circ F} & \mathcal{D'}\circ F & \imath _{\mathfrak{C'}\circ F} \ar@{=>}[r]_{\gamma ' \circ F} & \mathcal{D'}\circ F \ar@{=>}[u]_{\xi}  }}
\end{equation}
\begin{example}[canonical morphisms]
 For any reconstruction context $\mathcal{D}$ there is a canonical
right morphism $I_{R}:\mathcal{D}\rightarrow\mathcal{I}$, defined
as follows. In applicable prestacks it is the inclusion functor, i.e,
$F=\imath_{\mathfrak{C}}$. Among deleting processes it is just the
transformation $\xi_{C}=\gamma_{C}$ of $\mathcal{D}$. The existence
of left morphisms $I_{L}:\mathcal{D}\rightarrow I$ is more restrictive:
if we keep the canonical choice $F=\imath_{\mathfrak{C}}$, the condition
$\xi_{C}\circ\gamma_{C}=id_{C}$ implies that $\xi$ is a retraction
for $\gamma$. In the general case, the commutativity condition is
$\xi\circ\gamma=id\circ F$, which is also some kind of retraction
requirement (let us say that $\gamma$ has $\xi$ as a \emph{$F$-retraction}).
So, we have the following proposition:
\end{example}
\begin{proposition}
Given a prestack $F:\mathfrak{C}\rightarrow[\mathbb{X}^{op};\mathbf{Cat}]$,
there is a left morphism $D\rightarrow I$ of reconstruction contexts
coindicing with $F$ in applicable prestacks iff $\gamma$ has a $F$-retraction.
\end{proposition}
We could think of getting morphisms in the opposite direction, i.e,
from the trivial context $I$ to a given context $D$. This is an
even more strong requirement. This is essentially because a priori
we have no canonical functor $F:[\mathbb{X}^{op};\mathbf{Cat}]\rightarrow\mathfrak{C}$.
The commutativity conditions implies that if it exists, then it must
be a retraction for the inclusion of $\mathfrak{C}$ in $[\mathbb{X}^{op};\mathbf{Cat}]$.
In the case of left morphisms this is enough. For right morphisms
$I\rightarrow D$ we also need $\gamma$ to have a $F$-retraction.
\begin{example}[from RC to dRC]
 \label{morphism_RC_dRC}Let $\mathfrak{C}$ be the category of proper
stacks. Let $\Delta$ and $d$ be the contexts of the RC and dRC,
as defined in Example \ref{RC_dRC_contexts}. We have a right morphism
$F:d\rightarrow\Delta$, defined as follows. In applicable prestacks
we define $F:\mathfrak{C}\rightarrow[\mathbb{X}^{op};\mathbf{Cat}]$
as the inclusion of proper prestacks and $\xi:\Delta\Rightarrow d$
as the composition $\xi^{I}\circ u$ in diagram (\ref{dRC_diagram}).
Notice that the commutativity of (\ref{morphism_conjectures}) follows
directly from the commutativity of (\ref{dRC_diagram}).
\end{example}
\begin{proposition}
\label{prop_morphism_conjecture}Let $F:\mathcal{D}\rightarrow\mathcal{D}'$
be a morphism. For a given prestack $C\in\mathfrak{C}$:
\begin{enumerate}
\item if $F$ is left and $\mathcal{D}'$-RC-$F(C)$ holds, then $\mathcal{D}$-RC-$C$
holds;
\item if $F$ is right and $\mathcal{D}$-RC-$C$ holds, then $\mathcal{D}'$-RC-$F(C)$
holds.
\end{enumerate}
\end{proposition}
\begin{svmultproof}
For the first case, notice that the commutativity of (\ref{morphism_conjectures})
gives us $\xi_{C}\circ\gamma_{C}=\gamma'_{F(C)}$ and recall that
essentially injective functors behave as monomorphisms, so that if
$\gamma'_{F(C)}$ is essentially injective, then $\gamma_{C}$ is
too. But the validity of $\mathcal{D}'$-RC-$F(C)$ is, by definition,
the garantee that $\gamma'_{F(C)}$ is essentially injective. For
the second case, (\ref{morphism_conjectures}) gives $\gamma_{C}=\xi_{C}\circ\gamma'_{F(C)}$.
Now use the same argument of the first case.
\end{svmultproof}

In some cases we have a left morphism, we know that $\mathcal{D}$-RC-$C$
holds and we would like to conclude that $\mathcal{D}'$-RC-$F(C)$
holds. In other words, we would like to have conditions under which
the hypothesis of the first part of the last proposition implis the
conclusion of the second part, and vice-versa. This can be easily
ensured if we work with a special class of morphisms. We say that
a left morphism $F:\mathcal{D}\rightarrow\mathcal{D}'$ is a left
$C$\emph{-implication }if the transformation $\xi:D'\circ F\Rightarrow D$
is objectwise essentially injective. Right $C$-implications are defined
analogously.
\begin{proposition}
\label{prop_implication} Let $F:\mathcal{D}\rightarrow\mathcal{D}'$
be a morphism. For a given prestack $C\in\mathfrak{C}$:
\begin{enumerate}
\item if $F$ is left $C$-implication and $\mathcal{D}$-RC-$C$ holds,
then $\mathcal{D}'$-RC-$F(C)$ holds;
\item if $F$ is right $C$-implication and $\mathcal{D}'$-RC-$F(C)$ holds,
then $\mathcal{D}$-RC-$C$ holds.
\end{enumerate}
\end{proposition}
\begin{svmultproof}
As in the last proposition, for the first case (\ref{morphism_conjectures})
gives $\xi_{C}\circ\gamma_{C}=\gamma'_{F(C)}$. Since composition
of essentially injective functors remains essentially injective, it
is done. The second case is analogous.
\end{svmultproof}

\begin{corollary}
For any $\mathcal{D}$ and any $C\in\mathfrak{C}$, the conjecture
$\mathcal{D}$-RC-$C$ holds iff the canonical morphism $I_{R}:\mathcal{D}\rightarrow\mathcal{I}$
is a right $C$-implication.
\end{corollary}
\begin{svmultproof}
Straightforward.
\end{svmultproof}

As a final result, let us show that reconstruction conjectures are
invariant by a certain base-change.
\begin{proposition}
\label{base_change_conjectures}Let $\mathcal{D}$ be a reconstruction
context and suppose that $\mathcal{D}$-RC-$C$ holds for some applicable
prestack $C\in\mathfrak{C}$. In this case, if $f:A\rightarrow C$
is some objectwise essentially injective morphism in $\mathfrak{C}$,
then $\mathcal{D}$-RC-$A$ holds.
\end{proposition}
\begin{svmultproof}
Since $\gamma:\imath_{\mathfrak{C}}\Rightarrow D$ is a natural transformation,
we have $Df\circ\gamma_{A}=\gamma_{C}\circ f$. By hypothesis $f$
and $\gamma_{C}$ are essentially injective, so that $Df\circ\gamma_{A}$,
and therefore $\gamma_{A}$, is also.
\end{svmultproof}

\subsection{Reconstruction of Labeled Structured Hypergraphs}

In this subsection we will show that the RC is true for any prestack
$S_{\mathfrak{s}}^{\ell}$ of labeled $\mathfrak{s}$-structured hypergraphs.
As a consequence, since from Example \ref{ex_labeled_hypergraph}
and Example \ref{structured_presheaf_proper} this prestack is proper,
it will follow from Proposition \ref{prop_morphism_conjecture} and
Example \ref{morphism_RC_dRC} that dRC-$S_{\mathfrak{s}}^{\ell}$
holds.
\begin{theorem}
\label{thm_reconstrucao_feynman} For any functor of structures $\mathfrak{s}$,
the $S_{\mathfrak{s}}^{\ell}$-RC holds. In other words, $\Delta S_{\mathfrak{s},V}^{\ell}$
is essentially injective for every $V$.
\end{theorem}
\begin{svmultproof}
Notice that $\mathbf{S}_{\mathfrak{s},V}^{\ell}=\mathbf{S}_{\mathfrak{s},V}\cap\mathbf{Hyp}_{V}^{\ell}$.
Therefore, $\mathbf{S}_{\mathfrak{s},V}^{\ell}$ can be embbeded in
an essentially injective way in $\mathbf{Hyp}_{V}^{\ell}$. This can
also be verified explicitly by a counting process:$\underset{\underset{\,}{\;}}{\;}$

\noindent \emph{Addendum.} Regard $\mathfrak{s}$ as a bifunctor $\mathfrak{s}:\mathbb{N}\times\mathbb{N}\rightarrow\mathbf{Set}$
and let $N:\mathbb{X}^{op}\times\mathbb{N}\times\mathbb{N}\rightarrow\mathbf{Cat}\times\mathbf{Ab}$
be the functor given by $N(V,i,j)=\varnothing\times\mathbb{N}\simeq\mathbb{N}$
for every $V,i,j$. We have a natural transformation $k:S_{\mathfrak{s}}^{\ell}\times\mathfrak{s}\Rightarrow N$
such that for every $V,i,j$ the map $k_{i,j}^{V}:\mathbf{S}_{\mathfrak{s},V}^{\ell}\times s_{i,j}\rightarrow\mathbb{N}$
is the rule that to any labeled $\mathfrak{s}$-structured hypergraph
$(G,\Phi,\epsilon)$ and to any element $a_{ij}\in s_{i,j}$ it assigns
the cardinality of $\epsilon_{i,j}^{-1}(a_{ij})$. We have
\begin{equation}
(n-1)k_{ij}^{V}(G,a_{ij})=\sum_{x\in V}k_{ij}^{V-x}(G-x,a_{ij}),\label{counting}
\end{equation}
where $n=\vert V\vert$. Suppose now that $G,G'\in\mathbf{S}_{\mathfrak{s},V}^{\ell}$
are isomorphic only as labeled hypergraphs. From Remark \ref{isomorphism_labeled_graphs}
there is a unique isomorphism, determined by the labelings. This implies
that $G$ and $G'$ will have the same counting (\ref{counting}).
In particular, $k_{ij}^{V}(G',a_{ij})=k_{ij}^{V}(G,a_{ij})$, allowing
us to conclude that $\epsilon'_{i,j}=\epsilon_{i,j}$ for every $i,j$,
and therefore $\epsilon'=\epsilon$. Consequently, $G'$ and $G$
are also isomorphic as $\mathfrak{s}$-structured hypergraphs.$\underset{\underset{\,}{\;}}{\;}$

Returning to the proof, from Proposition \ref{base_change_conjectures}
we only need to prove RC for labeled hypergraphs, i.e, that $\Delta D_{V}^{\ell}:\mathbf{Hyp}_{V}^{\ell}\rightarrow\sqcap\mathbf{Hyp}_{V-x}^{\ell}$
is essentially injective for every $V$. So, let $(G,\varphi)$ and
$(G',\varphi')$ be two labeled hypergraphs and suppose that there
exists $f_{x}:(G-x,\varphi_{x})\simeq(G'-x,\varphi'_{x})$ for every
$x$. Since we are working with isomorphic labeled hypergraphs defined
over the same vertex set, Remark \ref{isomorphism_labeled_graphs}
allows us to assume that the isomorphism on the vertices is given
by the identity map , i.e, $(f_{x})_{1}=id$ for every $x$. Furthermore,
the bijection $(f_{x})_{j}:(E_{j})_{x}\rightarrow(E'_{j})_{x}$on
the $j$-edges must be given by $(\varphi_{j})_{x}^{-1}\circ(\varphi'_{j})_{x}^{-1}$,
so that it can be clearly extended to a bijection between $E_{j}$
and $E'_{j}$ preserving the labelings, which gives $G\simeq G'$.
\end{svmultproof}

\begin{corollary}
\label{reconstruction_labeled_structured} If $C\subset S_{\mathfrak{s}}^{\ell}$
is concretely proper and objectwise essentially injetive, then both
conjectures RC-$C$ and dRC-$C$ holds.
\end{corollary}
\begin{svmultproof}
It follows directly from Proposition \ref{base_change_conjectures}.
\end{svmultproof}

\section{Contexts for Functional Analysis \label{sec_functional_analysis}}

When working with functional analysis we are dealing with certain
classes of spaces, each one with an associated ``dual space'', and
for which we know how to take tensor products. This leads us to define
a \emph{context for functional analysis} (or simply a \emph{context})
as a monoidal category $(\mathbf{A},\otimes,1)$ endowed with a functor
$\cdot^{\vee}:\mathbf{A}^{op}\rightarrow\mathbf{A}$, assigning to
each object $U\in\mathbf{A}$ its \emph{dual} $U^{\vee}$, which are
compatible in the sense that we have a \emph{compatibility transformation}
$\cdot^{\vee}\otimes\cdot^{\vee}\Rightarrow(\cdot\otimes\cdot)^{\vee}$
and a distinguished morphism $i:1\rightarrow1^{\vee}$.

We define a \emph{morphism} between two contexts $(\mathbf{A},\otimes,\cdot^{\vee})$
and $(\mathbf{B},\circledast,\cdot^{*})$ as a functor $F:\mathbf{A}\rightarrow\mathbf{B}$
which weakly preserves tensor products and duals. In other words,
it is an oplax monoidal functor\footnote{Recall our convention that in an oplax monoidal functor the arrow
between the neutral objects remains in the correct direction.} together with a transformation $F(\cdot^{\vee})\Rightarrow F(\cdot)^{*}$
such that the diagram below commutes. We then have the category $\mathbf{Cnxt}$
of contexts and morphisms between them.$$
\xymatrix{\ar[d] F(U^{\vee}\otimes V^{\vee}) \ar[rr]  && F((U\otimes V)^{\vee}) \ar[d] \\
\ar[d] F(U^{\vee})\circledast F(V^{\vee})&& F(U\otimes V)^{*} \ar[d] & F(1_{\otimes}) \ar[r] & F(1^{\vee}_{\otimes})  \\
F(U)^{*}\circledast F(V)^{*} \ar[rr] && (F(U)\circledast F(V))^{*} & \ar[u] 1_{\circledast} \ar[r] & 1_{\circledast}^{*} \ar[u]   }
$$
\begin{remark}
In some cases, the compatibility transformation $\cdot^{\vee}\otimes\cdot^{\vee}\Rightarrow(\cdot\otimes\cdot)^{\vee}$
and the distinguished map $1\rightarrow1^{\vee}$ are isomorphisms.
In such cases we say that we have \emph{strong contexts.} With the
same notion of morphisms they define a full subcategory $\mathbf{SCnxt}\subset\mathbf{Cnxt}$.
\end{remark}
\begin{example}[linear algebra]
 \label{ex_linear_algebra}Given a field $\mathbb{K}$, the category
$\mathbf{Vec}_{\mathbb{K}}$ of $\mathbb{K}$-vector spaces defines
a context for functional analysis when endowed with the tensor product
monoidal structure $(\otimes_{\mathbb{K}},\mathbb{K})$, with $U^{\vee}$
being the linear dual $U^{*}$. This is \emph{not} a strong context,
since for arbitrary vector spaces the canonical transformation $U^{*}\otimes_{\mathbb{K}}V^{*}\rightarrow(U\otimes_{\mathbb{K}}V)^{*}$
is not an isomorphism. On the other hand, the full subcategory $\mathbf{FinVec}_{\mathbb{K}}$
of finite-dimensional $\mathbb{K}$-vector spaces is a strong context.
More generally, the transformation $U^{*}\otimes_{\mathbb{K}}V^{*}\rightarrow(U\otimes_{\mathbb{K}}V)^{*}$
exists but may not be an isomorphism for arbitrary $R$-modules; but
they are if we consider finitely generated projective $R$-modules
\citep{algebra_lang}. So, with analogous structure, $\mathbf{Mod}_{R}$
is a context and the category of finitely generated projective $R$-modules
is a strong context.
\begin{example}[locally convex]
 \label{ex_locally_convex} Now we can take $\mathbf{A}$ as some
category of topological real vector spaces and continuous linear maps
and think of defining $U\otimes V$ and $U^{\vee}$ as $U\otimes_{\mathbb{R}}V$
and $B(U;\mathbb{R})\subset U^{*}$, respectively, endowed with some
topology\footnote{Here, $B(U;V)$ denotes the space of bounded linear maps.}.
There are many possible choices of topology, of course. For locally
convex spaces (lcs), there are at least three canonical ways to topologize
$U\otimes V$: the projective, the injective and the inductive topologies.
The choice of each of them produce a symmetric monoidal category $(\mathbf{LCS},\otimes,\mathbb{R})$
\citep{nuclear_spaces_GROTHENDIECK}. Also, there are many topologies
in $B(U;\mathbb{R})$, such as the topology of pointwise convergence
and uniform convergence in bounded sets. With each of them, we get
a pseudocontext structure in $\mathbf{LCS}$.
\begin{example}[nuclear Fr\'{e}chet]
 \label{ex_nuc_frechet} When the spaces are nuclear, the injective
and the projective topologies coincide \citep{nuclear_spaces_GROTHENDIECK,nuclear_trevis,Costello},
so that we have a more canonical pseudocontext structure. In general,
none of the topologies in $U^{\vee}=B(U;\mathbb{R})$ will induce
isomorphisms $\cdot^{\vee}\otimes\cdot^{\vee}\simeq(\cdot\otimes\cdot)^{\vee}$,
but they exist if we restrict to the full subcategory $\mathbf{NucFrec}$
of nuclear Fr\'{e}chet spaces, showing that $(\mathbf{Frec},\otimes_{p},\cdot^{\vee})$
is a strong context \citep{Costello,nuclear_trevis}. Also, for Fr\'{e}chet
spaces the projective and the inductive topologies coincide \citep{topological_vector_spaces},
meaning that in $\mathbf{NucFrec}$ we have a canonical symmetric
monoidal structure.
\begin{example}[subcontext]
\label{ex_subcontext} Let $(\mathbf{A},\otimes,\cdot^{\vee})$ be
a context. We define a \emph{subcontext }as a full subcategory $\mathbf{C}\subset\mathbf{A}$
such that $1\in\mathbf{C}$ and which is closed under $\otimes$ and
$\cdot^{\vee}$, meaning that $U\otimes V$ and $U^{\vee}$ belongs
to $\mathbf{C}$ when $U,V\in\mathbf{C}$. It then follows that $(\mathbf{C},\otimes,\cdot^{\vee})$
is a context and that the inclusion functor $\imath:\mathbf{C}\hookrightarrow\mathbf{A}$
is a morphism of contexts. Furthermore, if $\mathbf{A}$ is strong,
then $\mathbf{C}$ is also (the reciprocal is false as the last example
shows). In particular, the full subcategories $\mathbf{Ban}$ of Banach
spaces and $\mathbf{Hilb}$ of Hilbert spaces are subcontexts of $\mathbf{Frec}$
and therefore define themselves strong contexts for functional analysis.
\begin{example}[categories with duals]
 \label{categories_duals} There are many flavors of monoidal categories
whose objects or morphisms have duals (in the monoidal sense), e.g,
autonomous categories, pivotal categories, spherical categories, spacial
categories, compact closed categories and dagger monoidal categories
(see \citep{survey_categories_duals} for a survey). All of them define
a version of strong contexts whose $\cdot^{\vee}$ is actually an
ana-functor. Via Tannaka duality, they can be characterized as the
representation category of certain monoid objects \citep{tannaka,higher_tannaka}.
\end{example}
\end{example}
\end{example}
\end{example}
\end{example}
We can consider monoidal categories $\mathbf{A}$ endowed with a functor
$\cdot^{\vee}:\mathbf{A}^{op}\rightarrow\mathbf{A}$ without any compatibility
condition. In this case we will say that we have a \emph{pseudocontext}.
A \emph{morphism} between two of them is just a lax monoidal functor
$F:\mathbf{A}\rightarrow\mathbf{B}$ together with a transformation
$F(\cdot^{\vee})\Rightarrow F(\cdot)^{*}$, giving us a category $\mathbf{PCntx}$
which contains $\mathbf{Cntx}$.

\subsection{Analytic Expressions}

From now on we will assume that the pseudocontexts $(\mathbf{A},\otimes,\cdot^{\vee})$
are such that:
\begin{enumerate}
\item the category $\mathbf{A}$ have countable limits, coproducts and cokernels;
\item the tensor product $\otimes$ and the functor of duals $\cdot^{\vee}$
preserve limits.
\end{enumerate}
$\quad\;\,$For physical interpretation we will also require that
they become endowed with an additional functor $K:\mathbf{A}\rightarrow\mathbf{Set}$
and for every $k\in\mathbb{Z}_{\geq0}$ a transformation $S_{k}\times K(U^{\otimes_{k}})\rightarrow K(U^{\otimes_{k}})$
which is objectwise a group action\footnote{Here $S_{k}$ is the permutation group and $U^{\otimes_{k}}=((U\otimes U)\otimes U)...$,
$k$ times.}. We require that the quotient $K(U^{\otimes_{k}})/S_{k}$ belongs
to the image of $K$ and have a single pre-image, so that it uniquelly
defines an object in $\mathbf{A}$. The obvious notation for this
object should be $\operatorname{Sym}^{k}(U)$, but here we will use
$P^{k}U$ to denote it. For $k>1$ we will say that this is the \emph{object
of $k$-propagators }in $U$ (2-propagators are called propagators
for short).

Given $X\in\mathbf{A}$ we define an \emph{object of formal power
series }as the countable product of copies of $X$, i.e, as the product
$\prod_{i}X_{i}$, with $X_{i}\simeq X$. We usually use the powers
of a formal parameter $h$ to indicate the order of the products,
writing $\prod_{i}X_{i}h^{i}$ or $X[[h]]$ for short. We can consider
power series not only with a single parameter $h$, but with a family
$\mathfrak{t}=(h_{n})_{n\in\mathbb{Z}_{\geq0}}$ of them. These will
be given by $X[[\mathfrak{h}]]=\prod_{n}\prod_{i}X_{i}^{n}h_{n}^{i}$
with $X_{i}^{n}\simeq X$. In physical contexts $\mathfrak{h}$ will
represent the family of fundamental parameters over which we will
do perturbation theory.

Write $PU=\prod_{k}P^{k}U$ and $\mathcal{O}U=PU^{\vee}$. Given a
family of parameters $\mathfrak{h}$ we define an \emph{interacting
term }in $U$ relative to $\mathfrak{h}$ as a morphism $I:1\rightarrow\mathcal{O}U[[\mathfrak{h}]]$.
A \emph{morphism} of interacting terms is a morphism in the under
category $1/\mathbf{A}$, giving us a full subcategory $\mathbf{Int}_{U,\mathfrak{h}}$
of $1/\mathbf{A}$.

Feynman rules will take structured hypergraphs and assigns to each
of them interacting terms and propagators which together will fit
into analytic expressions. Let $A^{k}U$ denote $(U^{\vee})^{\otimes_{k}}\otimes U^{\otimes_{k}}$
and let $AU=\prod_{k}A^{k}U$ (since the tensor product $\otimes$
preserves countable products, if $\mathbf{A}$ is a \emph{strong context},
we can also write $AU=PU\otimes PU^{\vee}$). An \emph{analytic expression
}in $U$ is a morphism $a:1\rightarrow AU$. A \emph{morphism }is
a morphism in $1/\mathbf{A}$, so that we have a full subcategory
$\mathbf{A}_{U}$. We say that an analytic expression\emph{ }has\emph{
order $k$ }if it takes values in $A^{k}U$ instead of $AU$, defining
a category $\mathbf{A}_{U}^{k}$. We have $\mathbf{A}_{U}\simeq\prod_{k}\mathbf{A}_{U}^{k}$.
\begin{proposition}
For every $U\in\mathbf{A}$, the category $\mathbf{A}_{U}$ acquires
a canonical monoidal structure, induced from the monoidal structure
in $\mathbf{A}$.
\end{proposition}
\begin{svmultproof}
Recall that if $(\mathbf{C},\otimes,1)$ is any monoidal category
and then for any comonoid object $X\in\mathbf{C}$, say with coproduct
$\mu:X\rightarrow X\otimes X$ and counit $\eta:X\rightarrow1$, then
the under category $X/\mathbf{C}$ can be endowed with a monoidal
product $X/\otimes$, defined by tensoring with $\otimes$ and precomposing
with $\mu$, whose neutral object is $\eta$. Particularly, for $X=1$
we get that $1/\mathbf{C}$ is monoidal. Let $(\mathbf{A},\otimes,\cdot^{\vee})$
be a context. Let $\mathbf{A}U\subset\mathbf{A}$ be the full subcategory
whose single object is $AU$. Since $\otimes$ and $\cdot^{\vee}$
preserve countable products, we see that $AU$ is actually a monoid
object, so that $\mathbf{A}U$ is a monoidal subcategory. On the other
hand, $\mathbf{A}_{U}\simeq1/\mathbf{A}U$, from where we get the
desired monoidal structure on analytic expressions.
\end{svmultproof}

\begin{proposition}
\label{functor_analytic_expressions} Every pseudocontext morphism
$F:\mathbf{A}\rightarrow\mathbf{B}$ induces, for each $U\in\mathbf{A}$,
a functor $F_{U}:\mathbf{A}_{U}\rightarrow\mathbf{B}_{F(U)}$ between
the corresponding categories of analytic expressions.
\end{proposition}
\begin{svmultproof}
Given an analytic expression $a:1\rightarrow AU$ in $\mathbf{A}$,
define $F_{U}(a)$ as the following composition. $$
\xymatrix{1_{\circledast} \ar[r] & F(1) \ar[r]^-{F(a)} & F(AU) \ar[r]^-u & \prod _k F(A^k U) \ar[r] & \prod _k B^k F(U) = BF(U) }
$$The first and the last arrows appear because $F$ is oplax monoidal
and because a pseudocontext morphism becomes endowed with a transformation
$F(\cdot^{\vee})\rightarrow F(\cdot)^{*}$. Furthermore, $u$ is from
the universality of products in $\mathbf{B}$. From the definition
of $F_{U}(a)$ it immediately follows that $F_{U}$ is functorial.
\end{svmultproof}

\begin{remark}
Unless the pseudocontext morphism $F:\mathbf{A}\rightarrow\mathbf{B}$
is a bilax monoidal functor (instead of only oplax monoidal), the
induced functor $F_{U}:\mathbf{A}_{U}\rightarrow\mathbf{B}_{F(U)}$
generally will not be monoidal. Indeed, recall that the monoidal structure
of $\mathbf{A}_{U}$ is essentially the monoid object structure of
$AU$ in $\mathbf{A}$. Therefore, saying that $F_{U}$ is monoidal
we are saying that $F$ maps the monoid $AU$ into the monoid $BF(U)$,
which is not a typical property of oplax functors, but which is clearly
satisfied when $F$ is bilax or strong monoidal.
\begin{remark}
We could think of replacing the products $\prod_{k}$ by coproducts
$\coprod_{k}$ in the above definitions and constructions. This would
make life easier. But, in order to do this, we should assume that
$\mathbf{A}$ has countable colimits (instead of countable limits)
and that $\otimes$ and $\cdot^{\vee}$ preserve colimits (instead
of limits). For our purposes, these assumptions are restrictive: if
$\otimes$ preserves colimits, then it has right adjoint, meaning
that $(\mathbf{A},\otimes,1)$ is a closed monoidal category and automatically
excluding $(\mathbf{Frec},\otimes_{p},\cdot^{\vee})$ as a possible
context.
\end{remark}
\end{remark}

\section{Feynman Functors\label{Feynman_functors}}

We can finally introduce and prove existence and uniqueness of functorial
Feynman functors. These assign to each presheaf of hypergraphs an
analytic expression in some pseudocontext. If the pseudocontext is
sufficiently well behaved, then it will be able to evaluate these
analytic expressions, giving us some kind of ``amplitude of probability''
assigned to each hypergraph.

Given a prestack $C:\mathbb{X}^{op}\rightarrow\mathbf{Cat}$ of hypergraphs
(generally regarded as an applicable prestack), a pseudocontext $(\mathbf{A},\otimes,\cdot^{\vee})$
and a functor $\tau:\mathbf{FinSet}\rightarrow\mathbf{A}$, we define
a \emph{functorial Feynman functor }(or \emph{Feynman functor})\emph{
}for $C$ with values in $(\mathbf{A},\tau)$, denoted by $Z:C\rightarrow(\mathbf{A},\tau)$,
as a function that to each $V\in\mathbf{FinSet}$ assigns a functor
$Z_{V}:C_{V}\rightarrow1/\mathbf{A}$ such that for every $G\in C_{V}$
we have $Z_{V}(G)\in\mathbf{A}_{\tau(VG)}$, where $VG$ is the vertex
set of $G$. We say that a Feynman functor is \emph{complete} when
each $Z_{V}$ is essentially injective, meaning that the structured
hypergraphs can be totally described by their associated analytic
expressions.

We are also interested in \emph{monoidal Feynman functors}. In order
to define them we need to work with $C$ proper, which means that
it becomes endowed with a transformation $\xi_{V,V'}:C_{V}\times C_{V'}\rightarrow C_{V\sqcup V'}$,
as discussed in Remark \ref{proper_prestack}. From it we obtain the
following transformation, which takes into account three finite sets
instead of only two. Notice that $\xi_{(V,V'),V''}\simeq\xi_{V,(V',V'')}$,
where the isomorphism means that any three hypergraphs have isomorphic
image under these maps.\begin{equation}{ \label{diagrama_xi}
\xymatrix{\ar@{-->}[rrd]_{\xi _{(V,V'),V''}} (C_V \times C_{V'}) \times C_{V''} \ar[rr]^-{\xi _{V,V'} \times id_{V''}} && C_{V\sqcup V'} \times C_{V''} \ar[d]^{\xi _{(V\sqcup V'),V''}} \\
&& C_{(V\sqcup V') \sqcup V''}   }}
\end{equation}

We say that $Z$ is \emph{oplax monoidal} if for every $V,V'\in\mathbf{FinSet}$,
every $G\in C_{V}$ and every $G'\in C_{V'}$ we have morphisms 
\begin{equation}
\mu_{V,V'}:Z_{V\sqcup V'}(\xi_{V,V'}(G,G'))\rightarrow Z_{V}(G)\otimes Z_{V'}(G')\quad\text{and}\quad\nu:Z_{\varnothing}(\varnothing)\rightarrow id_{1}\label{monoidal_feynman_rules}
\end{equation}
in $1/\mathbf{A}$ satisfying the usual comonoid-like diagrams (e.g,
the associativity diagram is that presented below\footnote{In order to simplify the notation we wrote $Z_{V,V'}$ instead of
$Z_{V\sqcup V'}$. Furthermore, notice that it makes sense due to
the commutativity of (\ref{diagrama_xi}).}). Similarly, we define the situations when $Z$ is \emph{lax monoidal}
and \emph{strong monoidal} (or simply \emph{monoidal}) by reverting
the arrow or by requiring that they are isomorphisms, respectively.
\begin{equation}{\label{oplax_feynman_functor}
\xymatrix{\ar[d]_{\mu _{V,(V',V'')}} Z_{V, (V',V'')}(\xi _{V, (V',V'')}(G,G',G'')) \ar[r]^{\simeq} & Z_{(V, V'),V''}(\xi _{(V, V'),V'')}(G,G',G'')) \ar[d]^{\mu _{(V,V'),V''}} \\
 \ar[d]_{id \times \mu _{V',V''}} Z_V (G) \otimes Z_{V',V''}(\xi _{V',V''}(G',G'')) & Z_{V,V'}(\xi _{V,V'}(G,G')) \otimes Z_{V''} (G'') \ar[d]^{\mu _{V,V'} \times id} \\
  Z_V (G) \otimes ( Z_{V'} (G') \otimes  Z_{V''} (G'')) \ar[r]_{\simeq} &  (Z_V (G) \otimes  Z_{V'} (G')) \otimes  Z_{V''} (G'') }}
\end{equation}
\begin{remark}
When $C$ is concretely proper, $C_{V}\subset\mathbf{Hyp}_{V}$ for
every $V$, so that 
\[
Z_{V\sqcup V'}(\xi(G,G'))\in\mathbf{A}_{\tau(V\sqcup V')}\quad\text{and}\quad Z_{V}(G)\otimes Z_{V'}(G')\in\mathbf{A}_{\tau V}\otimes\mathbf{A}_{\tau V'}.
\]
Therefore, in order to formalize the notion of lax (resp. oplax) monoidal
Feynman functors, instead of requiring the morphisms (\ref{monoidal_feynman_rules})
to belong to $1/\mathbf{A}$, in these cases we could require that
$\tau:(\mathbf{FinSet},\sqcup,\varnothing)\rightarrow(\mathbf{A},\otimes,1)$
is lax (resp. oplax) monoidal. However, this would produce a much
more rigid concept, e.g, $\tau$ could not be chosen as the constant
functor in some $U\in\mathbf{A}$, with $U\neq1$. And, as we will
see, it is precisely when $\tau=\operatorname{cst}$ that the explict
connection with perturbative quantum field theory is made.
\end{remark}
\begin{example}[trivial Feynman functor]
 For any $C$ and any $(\mathbf{A},\tau)$ there always exists a
trivial monoidal Feynman functor $cst_{1}:C\rightarrow(\mathbf{A},\tau)$
assigning to each $V$ the functor $cst_{1}:C_{V}\rightarrow1/\mathbf{A}$
constant in the neutral object $id_{1}$ of the monoidal category
$1/\mathbf{A}$. In the next section we will show that nontrivial
Feynman functors also exist.
\begin{example}[Feynman subfunctor]
\label{feynman_subfunctor} Suppose given a Feynman functor $Z:C\rightarrow(\mathbf{A},\tau)$.
Precomposition defines a Feynman functor $Z':C'\rightarrow(\mathbf{A},\tau)$
for any subfunctor $C'\subset C$, and:
\end{example}
\begin{enumerate}
\item if $Z$ is lax monoidal and $C'\subset C$ is proper, then $Z'$ is
lax monoidal. The same holds for oplax or strong monoidal;
\item if $Z$ is complete and $C'\subset C$ is essentially injetive, then
$Z'$ is complete.
\end{enumerate}
\end{example}
Our aim in this section is to show that when fixed a decomposition
system in a prestack $C$, we can obtain Feynman functors $F:C\rightarrow(\mathbf{A},\tau)$
constructively by means of following certain rules, called \emph{Feynman
rules}. So, let us begin by defining what we mean by a decomposition
system.

Let $\mathbf{C}\subset\mathbf{Hyp}$ be a category of hypergraphs.
A \emph{decomposition system} \emph{of order $n$} in $\mathbf{C}$
is a rule $\mathfrak{d}$ assigning to each $G\in\mathbf{C}$ decompositions
$EG_{j}=EG_{j}^{0}\sqcup EG_{j}^{1}\sqcup...\sqcup EG_{j}^{n}$, for
each $j\geq1$. In other words, we have a decomposition of the vertex
set $VG=EG_{1}$ and of each set $EG_{j}$ of $j$-edges into $n$
parts. We do not require conditions on the adjacency functions $\psi_{j}$.
We say that $v^{0}\in EG_{1}^{0}$ and $e_{j}^{0}\in EG_{j>1}^{0}$
are the \emph{external vertices} and \emph{external $j$-edges}, respectively.
If $i>0$, the elements $v^{i}$ of $EG_{1}^{i}$ and $e_{j}^{i}$
of $EG_{j}^{i}$ are called \emph{internal vertices }and \emph{internal}
$j$\emph{-edges of} \emph{type $i$}.

We say that $\mathfrak{d}$ is \emph{functorial} if each hypergraph
morphism $f:G\rightarrow G'$ preserve the decomposition. It is straightforward
to check that the choice of a functorial decomposition system induces
an embedding $C_{V}\hookrightarrow\mathbf{S}_{\mathfrak{s}}$ for
certain $\mathfrak{s}$. We define a functorial decomposition system
$\mathfrak{d}$ for a prestack $C$ as a rule that to any $V\subset X$
it assigns a functorial decomposition system $\mathfrak{d}_{V}$ for
$C_{V}$.
\begin{example}[trivial]
 Each $\mathbf{C}\subset\mathbf{Hyp}$ possesses a trivial functorial
decomposition system of arbitrary order: just define $VG^{0}=VG$,
$EG_{j>1}^{0}=EG_{j>1}$ and the remaining pieces given by the empty
set, i.e, $EG_{j}^{i}=\varnothing$ if $i>0$.
\begin{example}[incidence]
 \label{incidence}On the other hand, each $\mathbf{C}$ also possesses
a nontrivial functorial decomposition of order 2. In fact, recall
that to each hypergraph $G$ we assign a bipartite graph $IG$: its
\emph{incidence graph}, whose set of vertices is $V_{IG}=VG\sqcup EG$,
where $EG=\sqcup_{j}EG_{j}$. There exists a 2-edge between $x,x'\in V_{IG}$
iff $x\in VG$ and if $x'\in EG$ is some hyperedge adjacent to $x$.
The rule $G\mapsto IG$ is functorial and actually an equivalence
$\mathbf{Hyp}\simeq\mathbf{BipGrph}$ between hypergraphs and bipartite
graphs \citep{hypergraph_theory}. In particular, each $\mathbf{C}\subset\mathbf{Hyp}$
is equivalent to some category of bipartite graphs. But bipartite
graphs have, by definition, a decomposition of order $2$. So, any
$\mathbf{C}$ has an induced decomposition, which we denote by $\mathfrak{d}_{\mathfrak{i}}$.
Furthermore, each prestack $C$ can be endowed with this decomposition.
Particicularly, each $C$ can be embedded into a structured prestack
$S_{\mathfrak{s}}$ for certain $\mathfrak{s}$.
\begin{example}[Feynman]
\label{feynman_decomposition} For any $V$, the category $\mathbf{Fyn}_{V}$
(and, therefore the prestack) of Feynman graphs have a canonical functorial
decomposition system of order $2$, as explained in Example \ref{ex_feynman_graph}.
Analogously for Feynman hypergraphs and generalized Feynman hypergraphs
introduced in Example \ref{ex_struct_Feynman}.
\begin{example}[structured]
\label{structured_decomposition} Any prestack $S_{\mathfrak{s}}$
of additively structured hypergraphs (e.g, those which are finitely
structured - see Remark \ref{remark_additive_structures}) can be
endowed with a nontrivial functorial decomposition system of order
$2$. Indeed, define $VG^{0}$ as the subset of $VG$ consisting of
all vertices $v$ such that
\end{example}
\end{example}
\end{example}
\begin{enumerate}
\item[s1)] $d_{2}(v)=1$ and $d_{j>2}(v)=0$;
\item[s2)] for every $k$, $\mathfrak{s}_{k,1}(v)=0$;
\item[s3)] if exists $j$-edge between $v$ and $v'$, then $v'$ is not in
$VG^{0}$.
\end{enumerate}
In other words, $VG^{0}$ is the collection of vertices that have
trivial structure and that belong to a single $2$-edge. Then take
$VG^{1}=VG-VG^{0}$. Conditions s1) and s2) establish a subset $EG_{2}^{0}\subset EG_{2}$
and conditions s2) and s3) give us $\vert EG_{2}^{0}\vert=\vert VG^{0}\vert$.
We then have a decomposition $EG_{2}=EG_{2}^{0}\sqcup EG_{2}^{1}$.
Finally, for $j>2$ define $EG_{j}^{1}=EG_{j}$ and $EG_{j}^{i\neq1}=\varnothing$.
We will denote this decomposition system by $\mathfrak{d}_{\mathfrak{s}}$.

\end{example}
\begin{remark}
\label{standard_decomposition} For future reference, we define a
\emph{standard decomposition }$\mathfrak{d}$ for a prestack as being
one codifying the fundamental property of Examples \ref{feynman_decomposition}
and \ref{structured_decomposition}
\begin{enumerate}
\item[st1)]  \label{st1} the number of external vertices is equal to the number
of external edges. In other words, $\vert VG^{0}\vert=\sum_{j}\vert EG_{j}^{0}\vert$;
\item[st2)]  each $j$-edge in $EG_{j}^{0}$ is adjacent to at least one vertex
in $VG^{0}$.
\end{enumerate}
\begin{remark}
\label{normal_decomposition} In our construction we will need to
work with decompositions such that $\vert EG_{j}^{0}\vert\leq1$,
which cannot be satisfied by standard decompositions of order 2 with
$\vert VG^{0}\vert>1$. This leads us to consider decompositions such
that $\vert EG_{j}^{0}\vert\leq1$ and $\vert VG^{0}\vert\leq1$.
They will be called \emph{normal}. If $\mathfrak{d}$ is a functorial
normal decomposition of a prestack $C$ we say that the pair $(C,\mathfrak{d})$
is \emph{normal} and that $\mathfrak{d}$ is a \emph{normal structure}
\emph{for $C$}. Starting with any pair $(C,\mathfrak{d})$, where
$\mathfrak{d}$ is of order $n$, we can always build a normal pair
$(\mathcal{N}C,\mathcal{N}\mathfrak{d})$, with $\mathcal{N}\mathfrak{d}$
also of order $n$, called the\emph{ normalization }of $(C,\mathfrak{d})$
and defined as follows: $\mathcal{N}C_{V}$ is the category of hypergraphs
$\mathcal{N}_{\mathfrak{d}}G$ obtained from hypergrahs $G\in C_{V}$
by deleting $VG^{0}$ and $EG_{j}^{0}$, and consider the decompositon
$\mathcal{N}\mathfrak{d}_{V}$ given by 
\[
V\mathcal{N}_{\mathfrak{d}}G^{0}=\varnothing=E\mathcal{N}_{\mathfrak{d}}G_{j}^{0},\quad V\mathcal{N}_{\mathfrak{d}}G^{i>0}=VG^{i>0},\quad\text{and}\quad E\mathcal{N}_{\mathfrak{d}}G_{j}^{i>0}=EG_{j}^{i>0}.
\]
Furthermore, any prestack $C$ admits a normal structure $\mathfrak{d}_{0}$
of order 2: just take $VG^{0}=\varphi=EG_{j}^{0}$ and $VG^{1}=VG$
and $EG_{j}^{1}=EG_{j}$.
\begin{remark}
From Example \ref{incidence} any prestack $C$ admits a decomposition
$\mathfrak{d}_{\mathfrak{i}}$, implying $C\subset S_{\mathfrak{s}}$.
By Example \ref{structured_decomposition} $C$ can then be endowed
with a new decomposition $\mathfrak{d}_{\mathfrak{s}}$, which is
standard by the last remark. In other words, each $C$ admits a nontrivial
standard decomposition.
\end{remark}
\end{remark}
\end{remark}
Let $(C,\mathfrak{d})$ be a prestack of hypergraphs endowed with
a (non necessarily functorial) decomposition system $\mathfrak{d}$,
say of order $n$, let $(\mathbf{A},\otimes,\cdot^{\vee})$ be a pseudocontext
and let $\tau:\mathbf{FinSet}\rightarrow\mathbf{A}$ be a functor.
A \emph{Feynman rule} in $(C,\mathfrak{d})$ with coefficients in
$(\mathbf{A},\tau)$ is a rule $FR$ that to each $V\in\mathbf{FinSet}$
and each hypergraph $G\in C_{V}$ it assigns:
\begin{enumerate}
\item functions $r^{i}:VG^{i}\rightarrow\mathbb{Z}_{\geq0}$ and $r_{j}^{i}:EG_{j}^{i}\rightarrow\mathbb{Z}_{\geq0}$,
with $0\leq i\leq n$ and $j\geq1$, called the \emph{degrees }or
\emph{weights }of the decomposition $\mathfrak{d}$;
\item functors assigning to each external vertice, etc., a tensor in $\tau VG\in\mathbf{A}$
or $(\tau VG)^{\vee}\in\mathbf{A}$ of corresponding degree. Precisely,
functors
\[
FR^{i}:VG^{i}\rightarrow1/(\tau(VG)^{\vee}){}^{\otimes_{r^{i}}}\quad\text{and}\quad FR_{j}^{i}:EG_{j}^{i}\rightarrow1/(\tau VG)^{\otimes_{r_{j}^{i}}},
\]
regarding $VG^{i}$ and $EG_{j}^{i}$ as discrete categories, while
\[
1/(\tau(VG)^{\vee}){}^{\otimes_{r^{i}}}\quad\text{and}\quad1/(\tau VG)^{\otimes_{r_{j}^{i}}}
\]
denotes the full subcategories of $1/\mathbf{A}$ consisting of all
morphisms 
\[
1\rightarrow1/(\tau(VG)^{\vee}){}^{\otimes_{r^{i}(v^{i})}}\quad\text{and}\quad1/(\tau VG)^{\otimes_{r_{j}^{i}(e_{j}^{i})}},
\]
for every $v^{i}\in VG^{i}$ and $e_{j}^{i}\in EG_{j}^{i}$, respectively.
Recall that $1/U^{0}$ is the category with a single object $id_{1}$.
\end{enumerate}
\begin{example}[classic Feynman rules]
 \label{classical_feynman_rules} Let $C$ be any prestack endowed
with a functorial decomposition $\mathfrak{d}$. Let $(\mathbf{A},\otimes,\cdot^{\vee})$
be any pseudocontext endowed with any functor $\tau:\mathbf{FinSet}\rightarrow\mathbf{A}$.
An usual setup for Feynman rules $FR:(C,\mathfrak{d})\rightarrow(\mathbf{A},\tau)$
is to take for each $G\in C_{V}$ the degree functions as $r^{0}\equiv0$,
$r_{j}^{0}\equiv1$, $r_{j}^{i\geq0}\equiv j$ and $r^{i\geq0}(v)=d(v)=\sum_{j}d_{j}(v)$,
so that 
\[
FR^{0}:VG^{0}\rightarrow1/(\tau(VG)^{\vee})^{0}
\]
must be the constant functor in $id_{1}$ (i.e, it must assign the
trivial tensor in $\tau(VG)$ to each external vertice). Furthermore,
to each hypergraph $G$ consider two formal parameters $t,o$ such
that if $G$ is $b$-bounded and $\mathfrak{d}$ has order $n$, then
$t^{j}=0$ for $j>b$ and $u^{i}=0$ for $i>n$. Take the object $\tau(VG)[[t,o]]$
of formal power series in $t$, which decomposes as a product $\prod_{j,i}X_{i,j}o^{i}t^{j}$,
with $X_{i,j}\simeq\tau(VG)$. Then define $FR_{j}^{0}$ as any rule
assigning a generalized element $FR_{j}^{0}(e)$ of degree $(0,j)$
in $\tau(VG)$, i.e, as a morphism $1\rightarrow\tau(VG)[[t,o]]$
which actually take values in $X_{0,j}$. Moreover, define $FR_{j}^{i}$
as any map assigning $j$-propagators in $\tau(VG)$ to internal $j$-edges.
More precisely, recall that $P\tau(VG)[[t,o]]=\prod_{k,i,j}P_{i,j}^{k}o^{i}t^{j}$,
with $P_{i,j}^{k}\simeq\operatorname{Sym}^{k}(\tau(VG))$, and take
$FR_{j}^{i}(e)\in1/P_{i,j}^{j}$. Finally, take $FR^{i\geq0}$ as
any rule assigning to each internal vertex $v^{i}$ a corresponding
interacting term in $\tau(VG)^{\vee}$, relative to $t$, of degree
$d(v)$. In other words, for each $v$ we have a morphism $FR^{i}(v):1\rightarrow\mathcal{O}\tau(VG)[[t]]$,
where $\mathcal{O}\tau(VG)[[t,o]]=\prod_{k,i,j}I_{i,j}^{k}o^{i}t^{j}$,
with $I_{i,j}^{k}\simeq\operatorname{Sym}^{k}(\tau(VG)^{\vee})$,
which actually take values in $I_{i,1}^{d(v)}$.
\begin{example}[structured Feynman rules]
 \label{concrete_feynman_rules} We can particularize and consider
$C$ as a prestack $S_{\mathfrak{s}}$ of $\mathfrak{s}$-structured
hypergraphs. Furthermore, we can take $\tau:\mathbf{FinSet}\rightarrow\mathbf{A}$
as constant in some object $U$, called \emph{background object},\emph{
}so that all data assigned by $FR$ are tensors on $U$. We can also
use $\mathfrak{s}$ to require more properties on $FR^{i\geq0}$,
as follows. To each $s_{k,j}$ we consider a new formal parameter
$h_{k,j}$, so that 
\[
\mathcal{O}\tau(VG)[[\mathfrak{h}]]=\prod_{k,i,j,l_{1,1},l_{1,2},...}I_{i,j,l_{1,1},l_{2,1},...}^{k}t^{j}o^{i}h_{1,1}^{l_{1,1}}h_{1,2}^{l_{1,2}}...
\]
Then, for a given $(G,\epsilon)$ a $\mathfrak{s}$-structure graph,
take $FR^{i}(v)$ as a generalized element in $I_{i,1,\epsilon_{1,1}(v),\epsilon_{2,1}(v),...}^{d(v_{i})}.$
\begin{example}[physicists Feynman rules]
 \label{physics_feynman_rule} Let us now particularize even more
and consider $C$ as the concrete prestack $C_{V}=\mathbf{Fyn}_{V}$
of Feynman graphs of Example \ref{ex_feynman_graph} endowed with
the functorial decomposition of Example \ref{feynman_decomposition},
so that $b=2$ and $n=1$. Also, we have two nontrivial structures
$s_{1,1}=\mathbb{Z}_{2}$ and $s_{2,1}=\mathbb{Z}_{\geq0}$, corresponding
to the decomposition and the genus map, so that we have two parameters
$h_{1,1}$ and $h_{2,1}$. In this particular setting, let us denote
$h_{2,1}\equiv\hbar$. Furthermore, let us consider the pseudocontext
of Example \ref{ex_locally_convex}, allowing us to work with $\mathbf{LCS}$
instead of with $\mathbb{R}/\mathbf{LCS}$, as explained in Example
\ref{ex_Hahn_Banach}. Thus, according to the previous example, fixed
a background object $U$, we get Feynman rules by giving functions
$FR_{2}^{0}:EG^{0}\rightarrow U$, $FR_{2}^{1}:EG^{1}\rightarrow P^{2}U$
and $FR^{1}:VG^{1}\rightarrow\mathcal{O}U[\hbar]$ such that $FR^{1}(v)\in I_{g(v)}^{d(v)}$.
In turn, these functions are exactly the data defining the Feynman
rule of a perturbative QFT as will be more detailed explored in Section
\ref{applications}.
\end{example}
\end{example}
\end{example}
Suppose now that the decomposition system $\mathfrak{d}$ in $C$
is functorial. In this case, for any Feynman rule $FR$ in $(C,\mathfrak{d})$,
given a morphism $f:G\rightarrow G'$ we get the induced diagrams
below (without the segmented arrow). We say $FR$ is \emph{functorial}
when for each $f$ there exists the dotted arrows $r^{i}f$ and $r_{j}^{i}f$
completing these diagrams (of functors between discrete categories)
in a commutative way. As we will see, if we assume Axiom of Choice,
any Feynman rule becomes functorial. \begin{equation}{\label{functorial_feynman_rule}
\xymatrix{\ar[d]_{f^i} VG^{i} \ar[r]^-{FR^i} & 1/(\tau VG ^{\vee})^{\otimes_{r^{i}}} \ar@{-->}[d]^{r^{i}f} & \ar[d]_{f^i _j} EG_{j}^{i} \ar[r]^-{FR_j^i} & \ar@{-->}[d]^{r _{j}^{i}f} 1/(\tau VG)^{\otimes_ {r_{j}^{i}}} \\
VG\mathrm{'} ^{i} \ar[r]_-{FR\mathrm{'}^i} & 1/(\tau VG'^{\vee})^{\otimes _{r\mathrm{'}^{i}}} & EG\mathrm{'} _{j}^{i} \ar[r]_-{FR_{j}^i} &  1/(\tau VG')^{\otimes_{r\mathrm{'}_{j}^{i}}}  }}
\end{equation}

Let $\mathbf{A}$ be a category with a distinguished object $1\in\mathbf{A}$.
Let $\mathbb{D}$ be a collection of categories. We say that $(\mathbf{A},1)$
has the \emph{Hahn-Banach property} (or simply that it is Hahn-Banach)
\emph{relative to }$\mathbb{D}$ when for any $\mathbf{D},\mathbf{D}'\in\mathbb{D}$,
any full subcategoriesb$\mathbf{C},\mathbf{C}'\subset1/\mathbf{A}$
and any functors $F:\mathbf{D}\rightarrow\mathbf{D}'$, $G:\mathbf{D}'\rightarrow\mathbf{C}'$
and $H:\mathbf{D}\rightarrow\mathbf{C}$, the corresponding $e$ admit
an extension in $\mathbf{Cat}$ relatively to any functor $H$, as
in diagram (\ref{hahn_banach}). There, $e\circ m$ denotes the mono-epi
decomposition of $G\circ F$ obtained by taking its image in $\mathbf{Cat}$.
We say that $(\mathbf{A},1)$ is \emph{Hahn-Banach relative to a functor}
$\sigma:\mathbf{FinSet}\rightarrow\text{\ensuremath{\mathbf{Cat}}}$
if it is Hahn-Banach relative to $\mathbb{D}=\operatorname{img}(\sigma)$.
\begin{equation}{\label{hahn_banach}
\xymatrix{\ar[dd]_F \ar[rd]^e \mathbf{D} \ar[r]^H & \mathbf{C} \ar@{-->}[d]^{\hat{e}} \\
& \operatorname{img}(G\circ F) \ar[d]^m \\
\mathbf{D}' \ar[r]_G & \mathbf{C}'}}
\end{equation}
\begin{example}[locally convex spaces]
 \label{ex_Hahn_Banach} Let $\mathbf{LCS}$ be the category of lcs
endowed with the real line $\mathbb{R}$ as a distinguished object.
We assert that it is Hahn-Banach relative to the subcategory $\mathbf{FinVec}_{\mathbb{R}}$
of finite-dimensional vector spaces. With the inductive topology in
$\otimes$, $(\mathbf{LCS},\otimes,\mathbb{R})$ is closed monoidal,
whose internal hom $[U;U']$ between $U$ and $U'$ is given by $B(U;U')$
endowed with the topology of pointwise convergence. Therefore, $[\mathbb{R};U]\simeq U$
and we can work directly with $\mathbf{LCS}$ instead of $\mathbb{R}/\mathbf{LCS}$.
Since a continuous linear map taking values in a finite-dimensional
vector space is the same thing as a finite combination of linear functionals,
the Hahn-Banach condition above is a direct consequence of Hahn-Banach
theorem for lcs \citep{nuclear_trevis,topological_vector_spaces}.
The inclusion $\mathbf{Vec}_{\mathbb{R}}\hookrightarrow\mathbf{Set}$
is reflective. Let $\mathbb{R}[\cdot]:\mathbf{FinSet}\rightarrow\mathbf{FinVec}_{\mathbb{R}}$
be the restriction of the the left adjoint to the category of finite
sets and let $\imath:\mathbf{FinVec}_{\mathbb{R}}\hookrightarrow\mathbf{LCS}$
be the inclusion. We can verify that $\mathbf{LCS}$ is Hahn-Banach
relative to $\imath\circ\mathbb{R}[\cdot]$.
\begin{example}[full subcategories]
 If $(\mathbf{A},1)$ is Hahn-Banach, then $(\mathbf{C},1)$ is also
for every full subcategory $\mathbf{C}\subset\mathbf{A}$. So, in
particular, the pseudocontexts from Example \ref{ex_nuc_frechet}
and Example \ref{ex_subcontext} define Hahn-Banach pairs.
\end{example}
\end{example}
Now, the fundamental fact is that, as a consequence of the axiom of
choice, any pair $(\mathbf{A},1)$ is Hahn-Banach relative to the
functor regarding finite sets as discrete categories. Indeed, since
the domain category is discrete we can only work with objects, so
that the extension problem (\ref{hahn_banach}) is equivalent to a
problem is $\mathbf{Set}$ which has a solution, since $e$ is an
epimorphism and we are assuming Axiom of Choice.

Thus, if $FR:(C,\mathfrak{d})\rightarrow(\mathbf{A},\tau)$ is any
Feynman rule, the following dotted arrows exist, so that $FR$ is
a functorial Feynman rule for $r^{i}f=m^{i}\circ\hat{e}^{i}$ and
$r_{j}^{i}f=m_{j}^{i}\circ\hat{e}_{j}^{i}$.$$
\xymatrix{\ar[dd]_{f^i} \ar[rd]_-{e^i} VG^i \ar[r]^-{FR^i} & 1/(\tau(VG)^{\vee}){}^{\otimes_{r^{i}}} \ar@{-->}[d]^{\hat{e}^i} & \ar[dd]_{f^i_j} \ar[rd]_{e^i_j} EG^i_j \ar[r]^-{FR^i_j} & 1/(\tau(VG)^{\vee}){}^{\otimes_{r^i_j}} \ar@{-->}[d]^{\hat{e}}  \\
& \operatorname{img}(FR'^i\circ f^i) \ar[d]^{m^i} & & \operatorname{img}(FR'^i_j \circ f^i_j) \ar[d]^{m^i_j} \\
VG'^i \ar[r]_-{FR'^i} & 1/(\tau(VG')^{\vee}){}^{\otimes_{r^{i}}} & EG'^i_j \ar[r]_-{FR'^i_j} & 1/(\tau(VG')^{\vee}){}^{\otimes_{r^i_j}}}
$$

\subsection{Existence}

We can now prove that functorial Feynman rules induce monoidal Feynman
functors. Actually, we will need to work with Feynman rules $FR:(C,\mathfrak{d})\rightarrow(\mathbf{A},\tau)$
which are \emph{coherent} in the sense that the degrees $r^{i}$ and
$r_{j}^{i}$ of $\mathfrak{d}$ satisfy the following coherence equation:
\begin{equation}
\sum_{i\geq0}\sum_{j>2}\sum_{e_{j}^{i}}r_{j}^{i}(e_{j}^{i})=\sum_{i\geq0}\sum_{v_{i}}r^{i}(v_{i}).\label{coherence_equation}
\end{equation}

\begin{example}[standard Feynman rules]
 \label{standard_is_coherent} We say that a Feynman rule is \emph{standard
}if the decomposition system $\mathfrak{d}$ is standard (in the sense
of Remark \ref{standard_decomposition}) and if the degrees $r^{i}$
and $r_{j}^{i}$ are given by $r^{0}\equiv1$, $r^{i>0}(v^{i})=d_{i}(v^{i})$,
$r_{j}^{0}\equiv1$ and $r_{j}^{i>0}(e_{j}^{i})=j$. For instance,
the classic Feynman rules are standard. We assert if $FR$ is standard
then it is coherent. Due to the structure of the degrees, for any
hypergraph $G\in C_{V}$, the left-hand side and the right-hand side
(\ref{coherence_equation}) writes, respectively
\[
\sum_{j>2}\vert EG_{j}^{0}\vert+\sum_{i>0}\sum_{j>2}j\vert EG_{j}^{i}\vert\quad\text{and}\quad\vert VG^{0}\vert+\sum_{i>0}\sum_{v_{i}}r^{i}(v_{i}).
\]
Since $\mathfrak{d}$ is standard, the first term of both sides coincide.
Let $G_{i>0}$ be the hypergraph obtained by deleting $VG^{0}$. The
formula (\ref{degrees_vertices_edges_formula}) applied to it shows
that the second terms also coincide.
\end{example}
\begin{theorem}
\label{thm_feynman_rules} Let $(\mathbf{A},\otimes,\cdot^{\vee})$
be a pseudocontext endowed with a functor $\tau:\mathbf{FinSet}\rightarrow\mathbf{A}$
and let $C^{b}$ be a prestack of $b$-bounded hypergraphs. Suppose
one of the following conditions:
\begin{enumerate}
\item[\emph{c1)}]  $C^{b}\subset S^{\ell,b}$, i.e, our hypergraphs are labeled;
\item[\textit{\emph{c2)}}]  the monoidal category $(\mathbf{A},\otimes,1)$ is strict.
\end{enumerate}
Then any coherent Feynman rule $FR:(C^{b},\mathfrak{d})\rightarrow(\mathbf{A},\tau)$
defines a Feynman functor $Z:C^{b}\rightarrow(\mathbf{A},\tau)$.

\end{theorem}
\begin{svmultproof}
Assuming condition c1), let $(G,\varphi)$ be a labeled $b$-bounded
hypergraph in $C_{V}^{b}\subset S^{\ell,b}$. Let $n$ be the order
of $\mathfrak{d}$. The decompositions $VG=\sqcup_{i}V^{i}G$ and
$E_{j>1}G=\sqcup_{i}E_{j}^{i}G$ with $0\leq i\leq n$ defined by
$\mathfrak{d}$, induce decompositions for the bijections $\varphi_{j}$.
Without loss of generality we can assume that now we have $\varphi_{1}^{i}:[\vert V^{i}G\vert]\rightarrow V^{i}G$
and $\varphi_{j>1}^{i}$. Since morphisms of labeled hypergraphs preserve
the labelings, they will preserve these decomposed labelings. With
this in mind, if $FR:(C^{b},\mathfrak{d})\rightarrow(\mathbf{A},\tau)$
is a Feynman rule, for each set $V\subset X$ and each labeled hypergraph
$(G,\varphi)\in C_{V}$, define $Z_{V}^{i}(G)$ as the following tensor
product (in $1/\mathbf{A}$)
\begin{equation}
Z_{V}^{i}(G)=(...((FR^{i}(\varphi^{i}(1))\otimes FR^{i}(\varphi^{i}(2)))\otimes....)\otimes FR^{i}(\varphi^{i}(\vert V^{i}G\vert)).\label{labeled_feynman_functor-1}
\end{equation}
In analogous way, define $Z_{j,V}^{i}(G)$ for every $1<j\leq b$,
i.e, by taking the tensor product in $1/\mathbf{A}$ of the images
of $FR_{j}^{i}$, following the order given by the labelings $\varphi_{j}^{i}$.
Furthermore, define
\begin{equation}
Z_{V}^{I}(G)=(...((Z_{V}^{0}(G)\otimes Z_{V}^{1}(G))\otimes Z_{V}^{2}(G))...)\otimes Z_{V}^{n}(G)\label{feynman_functor_rules-1}
\end{equation}
and similarly $Z_{J,V}^{i}(G)$ for each $0<i\leq n$. Then take 
\begin{equation}
Z_{V}(G)=(...((Z_{V}^{I}(G)\otimes Z_{J,V}^{1}(G))\otimes Z_{J,V}^{2}(G))...)\otimes Z_{J,V}^{n}(G).\label{feynman_rules_2}
\end{equation}
From the coherence equation (\ref{coherence_equation}) we see that
$Z_{V}(G)\in\mathbf{A}_{\tau VG}$. Therefore, we have a rule $G\mapsto Z_{V}(G)$
assigning to each hypergraph with vertex set $VG$ an analytic expression
in $\tau VG\in\mathbf{A}$. Its functoriality follows from the functoriality
of the Feynman rules $FR$. Indeed, if $f:(G,\varphi)\rightarrow(G',\varphi')$
is a morphism in $C_{V}^{b}$, we have the morphisms $r^{i}f$ anf
$r_{j}^{i}$ in (\ref{functorial_feynman_rule}). Since $f$ preserves
the decomposed labelings, $r^{i}f$ induce morphisms $Z_{V}^{i}(f):Z_{V}^{i}(G)\rightarrow Z_{V}^{i}(G')$
and analogously for $r_{j}^{i}$. Define $Z_{V}^{I}(f)$ by replacing
the symbol ``$G$'' with the symbol ``$f$'' in (\ref{feynman_functor_rules-1})
and define $Z_{J,V}^{i}(f)$ similarly. Finally, replacing ``$G$''
with ``$f$'' in (\ref{feynman_rules_2}) we get the desired morphism
$Z_{V}(f):Z_{V}(G)\rightarrow Z_{V}(G')$. It is straightforward to
check that composition and identity morphisms are preserved, so that
we have a Feynman functor $Z:C\rightarrow(\mathbf{A},\tau)$. Finally,
notice that condition c1) was used only to fix an ordering in the
tensor products (\ref{labeled_feynman_functor-1}). Suppose now that
c2) is satisfied. Then $\mathbf{A}$ is strict, so that we can remove
the parenthesis of tensor products, meaning that we do not need to
care about ordering. So, the same construction works even without
labelings.
\end{svmultproof}

\begin{corollary}[of the proof]
 \label{corollary_feynman_functor}A Feynman functor induced by a
Feynman rule is never trivial unless the prestack $C^{b}$ is trivial.
\end{corollary}
\begin{svmultproof}
Just look at the defining expressions (\ref{labeled_feynman_functor-1}),
(\ref{feynman_functor_rules-1}) and (\ref{feynman_rules_2}) and
use the coherece equation.
\end{svmultproof}

We say that two Feynman rules $Z:C\rightarrow(\mathbf{A},\tau)$ and
$Z':C'\rightarrow(\mathbf{A}',\tau')$ \emph{differ by a change of
coefficients} if they are defined in the same prestack, i.e, if $C'=C$,
and if there exists an isomorphism $F:(\mathbf{A},\otimes,\cdot^{\vee})\rightarrow(\mathbf{A}',\otimes',\cdot^{\vee'})$
of pseudocontexts and a natural isomorphism $\zeta:\tau\simeq\tau'$
such that for every hypergraph $G\in C_{V}$ we have $F_{\zeta}\circ Z=Z'$,
where $F_{\zeta}:\mathbf{A}_{\tau VG}\simeq\mathbf{A}'_{\tau'VG}$
is the equivalence induced by $F$ and $\zeta$ (this equivalence
exists, since by Proposition \ref{functor_analytic_expressions} morphisms
of pseudocontexts induce functors between the categories of analytic
expressions).
\begin{proposition}
\label{general_existence} Up to change of coefficients, there exists
a nontrivial Feynman functor $Z:C\rightarrow(\mathbf{A},\tau)$ defined
in any nontrivial proper prestack $C$ (not necessarily bounded) and
taking values in any pseudocontext $(\mathbf{A},\otimes,\cdot^{\vee})$
endowed with any functor $\tau:\mathbf{FinSet}\rightarrow\mathbf{A}$.
\end{proposition}
\begin{svmultproof}
From Example \ref{incidence} any $C$ is subprestack of $S_{\mathfrak{s}}^{2}$
for some $\mathfrak{s}$. By Example \ref{structured_decomposition}
and by Remark \ref{standard_decomposition}, $S_{\mathfrak{s}}^{2}$
(and therefore $C$) can be endowed with a standard decomposition
system $\mathfrak{d}$, and by Example \ref{classical_feynman_rules}
we have a Feynman rule (the classical one) $FR:(S_{\mathfrak{s}}^{2},\mathfrak{d})\rightarrow(\mathbf{A},\tau)$
with coefficients in any $(\mathbf{A},\tau)$, which induces a Feynman
rule in $(C,\mathfrak{d})$. From Example \ref{standard_is_coherent}
this Feynman rule is coherent, so that if $\mathbf{A}$ is strict,
then by Theorem \ref{thm_feynman_rules} $FR$ induces a Feynman functor
$Z:C\rightarrow(\mathbf{A},\tau)$. But, since we are working up to
change of coefficients, Mac Lane's strictification theorem for monoidal
categories allows us to forget the strictness hypothesis on $(\mathbf{A},\otimes,1)$.
Nontrivialily of $Z$ is due Corollary \ref{corollary_feynman_functor}.
\end{svmultproof}

It would be interesting to find conditions on Feynman rules under
which the induced Feynman functors are monoidal and/or complete. We
start by analyzing completeness. Let $(\mathbf{C},\otimes,1)$ be
a monoidal category and let $F:\mathbf{I}\rightarrow\mathbf{C}$ be
some functor. We say that $F$ is \emph{decomposable }if for any isomorphism
$f:I\rightarrow I'$ in $\mathbf{I}$ and any decompositions $F(I)\simeq X\otimes Y$
and $F(I')\simeq X'\otimes Y'$ there are isomorphisms $\alpha:X\simeq X'$
and $\beta:Y\simeq Y'$ in $\mathbf{C}$ such that $F(f)=\alpha\otimes\beta$,
i.e, such that the diagram below commutes. If $(\mathbf{C},\otimes,\vee)$
is actually a pseudocontext, we require decompositions not only for
$F(I)$, but also for each $F(I)^{\otimes_{r}}$ and $(F(I)^{\vee})^{\otimes_{r}}$.
We say that $F$ is $1/$decomposable if the induced functor $1/F:\mathbf{I}\rightarrow1/\mathbf{C}$
is decomposable.$$
\xymatrix{\ar[d]_{F(f)} F(I) \ar[r]^{\simeq} & X\otimes Y \ar[d]^{\alpha \otimes \beta} \\
F(I') \ar[r]_{\simeq} & X' \otimes Y'  }
$$

We say that a Feynman rule $Z:(C,\mathfrak{d})\rightarrow(\mathbf{A},\tau)$
is \emph{complete} if for each hypergraph $G$ the maps $FR_{j}^{i}$
are equivalences over their images. This means that different internal
vertices and different external and internal hyperedges have different
tensorial representations. If, in addition, $FR^{i}$ are also equivalences
over their images, we say that $Z$ is \emph{strongly complete}.
\begin{proposition}
\label{existence_complete_feynman_functors} In the same notations
of Theorem \ref{thm_feynman_rules}, let $FR:(C^{b},\mathfrak{d})\rightarrow(\mathbf{A},\tau)$
be a coherent Feynman rule such that $\tau$ is $1/$decomposable
and suppose one of the following conditions:
\begin{enumerate}
\item[\emph{c1)}] $C^{b}\subset S^{b,\ell}$ is concrete and $FR$ is complete;
\item[\emph{c2)}] $C^{b}\subset S^{b,\ell}$ is $FR$ is strongly complete;
\item[\emph{c3)}] $(\mathbf{A},\otimes,1)$ is strict and $FR$ is strongly complete.
\end{enumerate}
Then $FR$ induces a Feynman functor which is also complete.

\end{proposition}
\begin{svmultproof}
It is clear that the induced monoidal Feynman functor exists, since
conditions c1), c2) and c3) above contains the conditions c1) and
c2) of Theorem \ref{thm_feynman_rules}. Let us show that the induced
Feynman functor is complete. We will work first with c1). Fixed $V$,
given a morphism $f:G\rightarrow G'$ in $C_{V}^{b}$, suppose that
$Z_{V}(f):Z_{V}(G)\rightarrow Z_{V}(G')$ is an isomorphism. From
the definition of $Z_{V}(f)$ and the fact that $\tau$ is $1/$decomposable
we find that $r^{i}f$ and $r_{j}^{i}$, with $i=0,1...,n$ and $1<j\leq b$,
are all isomorphisms. Because $FR$ is complete, each $FR_{j>1}^{i}$
is an equivalence over its image. From the commutativity of diagrams
(\ref{functorial_feynman_rule}) we then see that $f_{j}^{i}$ are
bijections. Since the prestack is concrete, we are working with hypergraphs
over the same vertex set. Furthermore, since the hypergraphs are labeled,
from Remark \ref{isomorphism_labeled_graphs} we can assume $f_{V}:V\rightarrow V$
equal to the identity $id_{V}$. Then each $f^{i}$ is also a bijection,
so that $f$ is a hypergraph isomorphism. Assume c2) or c3). Then
now $FR$ is strongly complete, so that not only $FR_{j>1}^{i}$,
but also $FR^{i}$, are equivalences over their images, implying directly
that $f_{j}^{i}$ and $f^{i}$ are bijections.
\end{svmultproof}

\begin{corollary}
\label{corollary_complete_functors} Let $(\mathbf{A},\otimes,\cdot^{\vee})$
be a pseudocontext and let $\tau$ $1/$decomposable. In this case:
\begin{enumerate}
\item[\emph{c1)}]  if $C\subset S_{\mathfrak{s}}^{\ell,b}$ is bounded, labeled, structured,
then for any normal structure $\mathfrak{d}$ in $C$ there exists
a nontrivial complete Feynman functor $Z:C\rightarrow(\mathbf{A},\tau)$;
\item[\emph{c2)}]  if $C$ is arbitrary, then for any normal structure $\mathfrak{d}$
in $C$ such a Feynman functor also exists, but only up to a change
of coefficients.
\end{enumerate}
\end{corollary}
\begin{svmultproof}
From Example \ref{classical_feynman_rules}, if $C\subset S_{\mathfrak{s}}^{b}$,
then there exist classic Feynman rules in $C$. Notice that whenever
classic rules of Feynman exist, we can modify them in order to assume
that $FR_{j}^{i>0}$ and $FR^{i>0}$ are equivalences over their images.
If $(C,\mathfrak{d})$ is normal, the domains of $FR^{0}$ and $FR_{j}^{0}$
are the empty category or the point category, so that both functors
are automatically equivalences over their images. Therefore, $FR$
becomes strongly complete. With this in mind, c1) and c2) follow,
respectivly, from conditions c2) and c3) of Proposition \ref{existence_complete_feynman_functors}.
\end{svmultproof}

Recall from Example \ref{normal_decomposition} that we have a rule
$\mathcal{N}$ assigning to each pair $(C,\mathfrak{d})$ its normalization
$(\mathcal{N}_{\mathfrak{d}}C,\mathcal{N}\mathfrak{d})$. We say that
some assertion about prestacks with decompositions holds \emph{up
to normalization} if it holds for every $\mathcal{N}_{\mathfrak{d}}C$.
\begin{corollary}
\label{corollary_normalization}Up to normalization and change of
coefficients, there exists a nontrivial complete Feynman functor defined
in any pair $(C,\mathfrak{d})$ and taking values in any pair $(\mathbf{A},\tau)$
with $\tau$ $1$/decomposable.
\end{corollary}
\begin{svmultproof}
Straightforward from condition c2) of Corollary \ref{corollary_complete_functors}.
\end{svmultproof}

\begin{remark}
The existence result for complete Feynman functors is much less general
than that for Feynman functors: for the complete case we need to work
with decomposable functors $\tau:\mathbf{FinSet}\rightarrow\mathbf{A}$.
This hypothesis on $\tau$ cannot be avoided because we have no analogue
of Mac Lane's strictification theorem establishing that any $(\mathbf{A},\tau)$
is equivalent to other $(\mathbf{A}',\tau')$ with a $\tau'$ $1/$decomposable.
\end{remark}
We will now discuss the monoidal property. Let $(C,\xi)$ be a proper
prestack endowed with a functorial decomposition $\mathfrak{d}$.
We say that a Feynman rule $FR:(C,\mathfrak{d})\rightarrow(\mathbf{A},\tau)$
is\emph{ oplax monoidal} if for any every $V,V'\subset X$ and for
every $G\in C_{V}$ and $G'\in C_{V'}$ we have functors 
\begin{eqnarray}
FR^{i}(G,G'):V\xi(G,G')^{i} & \rightarrow & FR^{i}(VG^{i})\otimes FR^{i}(VG'^{i})\label{feynman_rule_oplax_1}\\
FR_{j}^{i}(G,G'):E\xi(G,G')_{j}^{i} & \rightarrow & FR_{j}^{i}(EG_{j}^{i})\otimes FR_{j}^{i}(EG\mathrm{\mathrm{{'}}}_{j}^{i}),\label{feynman_rule_oplax_2}
\end{eqnarray}
and also $FR^{i}(\varnothing)\rightarrow id_{1}$ and $FR_{j}^{i}(\varnothing)\rightarrow id_{1}$,
fulfilling comonoid-like diagrams analogous to (\ref{oplax_feynman_functor}).
Lax monoidal and strong monoidal Feynman rules are defined in a similar
way.
\begin{example}[concrete proper]
\label{concret_proper} Let $(C,\xi)$ be a concrete proper prestack.
We say that a functorial decomposition $\mathfrak{d}$ in $C$ is
\emph{compatible} with $\xi$ if 
\begin{equation}
V\xi(G,G')^{i}=V\xi(G,\varnothing)^{i}\sqcup V\xi(\varnothing,G')^{i}\;\,\text{and}\;\,E\xi(G,G')_{j}^{i}=E\xi(G,\varnothing)_{j}^{i}\sqcup E\xi(\varnothing,G')_{j}^{i}\label{concrete_proper}
\end{equation}
for $0\leq i\leq n$ and $j>1$, where $n$ is the order of $\mathfrak{d}$.
For instance, if $C$ is concretely proper, which means that $\xi(G,G')=G\sqcup G'$,
then the condition above becomes 
\[
V(G\sqcup G')^{i}=VG^{i}\sqcup VG'^{i}\;\,\text{and}\;\,E(G\sqcup G')_{j}^{i}=EG_{j}^{i}\sqcup EG'{}_{j}^{i}.
\]
We notice that any Feynman rule $FR:(C,\mathfrak{d})\rightarrow(\mathbf{A},\tau)$,
defined in a concrete proper prestack endowed with a compatible decomposition,
is oplax monoidal in a unique way. Indeed, notice that since we have
(\ref{concrete_proper}), the maps (\ref{feynman_rule_oplax_1}) and
(\ref{feynman_rule_oplax_2}) we are looking for will be defined in
a coproduct. But maps defined in a coproduct are uniquely determined
by its components. So, $FR^{i}(G,G')$ exists and it is totally determined
by $FR^{i}(G)$ and $FR^{i}(G')$.
\end{example}
\begin{proposition}
\label{oplax_feynman_rule_functor} Under the same notations and hypotheses
of Theorem \ref{thm_feynman_rules}, if $C^{b}$ is proper and the
coherent Feynman rule is oplax (resp. lax or strong) monoidal, then
the induced Feynman functor is oplax (resp. lax or strong) monoidal.
\end{proposition}
\begin{svmultproof}
We will prove only the oplax case assuming condition c2) in Theorem
\ref{thm_feynman_rules}. The other cases (lax, strong and condition
c1) are analogous. So, let $FR:(C^{b},\mathfrak{d})\rightarrow(\mathbf{A},\tau)$
be a coherent Feynman rule with $\mathbf{A}$ strict and $FR$ oplax,
and let $Z$ be the induced Feynman rule. Notice that, by definition
$Z_{V,V'}(\xi(G,G'))$ is the tensor product between $Z^{i}(\xi(G,G'))$
and $Z_{j}^{i}(\xi(G,G'))$, for $0\leq i\leq n$, where here we omit
the subindices $V,V'$ in order to simplify the notation (see expressions
(\ref{feynman_rules_2}) and (\ref{feynman_functor_rules-1})). In
turn, $Z^{i}(\xi(G,G'))$ is the tensor product between $FR^{i}(v)$,
with $v\in V\xi(G,G')^{i}$, while $Z_{j}^{i}(\xi(G,G'))$ is the
tensor product of $FR_{j}^{i}(e)$, for $e\in E\xi(G,G')_{j}^{i}$.
Since $FR$ is oplax, we have the maps (\ref{feynman_rule_oplax_1}),
(\ref{feynman_rule_oplax_2}) and also $\nu:FR(\varnothing)\rightarrow id_{1}$.
Therefore, tensoring $FR^{i}(v)$ for every $v$ and $FR_{j}^{i}(e)$
for every $e$ we get maps $Z^{i}(\xi(G,G'))\rightarrow Z^{i}(G)\otimes Z^{i}(G')$
and $Z_{j}^{i}(G,G')\rightarrow Z_{j}^{i}(G)\otimes Z_{j}^{i}(G')$.
Tensoring them and varying $i$ and $j$, we obtain a map 
\[
\mu_{V,V'}:Z_{V,V'}(\xi(G,G'))\rightarrow Z_{V}(G)\otimes Z_{V'}(G').
\]
Furthermore, by definition of $Z$ we see that $\nu$ induces another
$\nu:Z_{\varnothing}(\varnothing)\rightarrow id_{1}$. These maps
will satisfy the comonoid-like diagrams precisely because they are
finite tensor products of maps that satisfy the diagrams.
\end{svmultproof}

\begin{corollary}
\label{existence_oplax}Up to change of coordinates, there exists
a nontrivial oplax monoidal Feynman functor defined in any concrete
proper prestack $C$ and taking values in any pseudocontex $\mathbf{A}$
endowed with any functor $\tau:\mathbf{FinSet}\rightarrow\mathbf{A}$.
\end{corollary}
\begin{svmultproof}
Direct consequence of Example \ref{concrete_feynman_rules}, Example
\ref{concret_proper} and Proposition \ref{oplax_feynman_rule_functor}.
\end{svmultproof}

\begin{remark}
We could think of getting a general existence result for Feynman functors
which are simultaneously complete and oplax monoidal. Corollary \ref{corollary_complete_functors}
tell us that complete Feynman functors exist with the hypothesis that
$C$ is endowed with a normal structure. This condition was avoided
in Corollary \ref{corollary_normalization} by making use of a normalization
process. Similarly, Corollary \ref{existence_oplax} states that oplax
monoidal Feynman functors exist if $C$ is concrete proper. So, we
could try to build some kind of ``concretization'' and ``propertification''
processes, allowing us to say that up to them oplax Feynman functors
always exist. The fundamental fact is that, even if these processes
are built, we \textbf{cannot} use them to say that \emph{up to normalization,
``propertification'' and ``concretification'' complete oplax Feynman
functor always exist}. This happens because the normalization of an
arbitrary nontrivial prestack cannot be concrete. Indeed, recall that
being concrete means that $C_{V}\in\mathbf{Hyp}_{V}$ for any $V$.
However, the normalization $\mathcal{N}C$ is such that $V\mathcal{N}_{\mathfrak{d}}G^{0}=\varnothing$,
so that $\mathcal{N}C_{V}\nsubseteq\mathbf{Hyp}_{V}$ unless the initial
decomposition $\mathfrak{d}$ coincide with $\mathcal{N}\mathfrak{d}$,
i.e, unless $(C,\mathfrak{d})$ is normal.
\end{remark}

\subsection{Uniqueness}

Closing our discussion on Feynman functors, let us focus on the uniqueness
problem. We will show that two complete Feynman functors are always
conjugated in a suitable way.

We say that a functor $\alpha:\mathbf{C}\rightarrow\mathbf{C'}$ is
\emph{quasi} \emph{essentially injective} (qei) if it is constant
in a subcategory $c\subset\mathbf{C}$ and essentially injective in
the remaining $\mathbf{C}-c$. We say that $F:\mathbf{C}\rightarrow\mathbf{D}$
is \emph{quasi essentially injectively conjugated }(qeic) to another
functor $F':\mathbf{C}'\rightarrow\mathbf{D}'$ if there are qei functors
$\alpha$ and $\beta$ making the following square commutative up
to natural isomorphisms:\begin{equation}{\label{quasi_essentially_injective}
\xymatrix{\ar[d]_{\alpha} \mathbf{C} \ar[r]^{F} & \mathbf{D} \ar[d]^{\beta} \\
\ar@{=>}[ru]^{\simeq} \mathbf{C'} \ar[r]_G & \mathbf{D'}}}
\end{equation}
\begin{lemma}
Let $\mathbf{D}$ a category with null objects. If a functor $F:\mathbf{C}\rightarrow\mathbf{D}$
is essentially injective and faithful, then it is qeic to any essentially
injective functor $G:\mathbf{C}\rightarrow\mathbf{D}$.
\end{lemma}
\begin{svmultproof}
Set $\alpha=id_{\mathbf{C}}$ and for every $Y\in\mathbf{D}$ define
\[
\beta(Y)=\begin{cases}
G(F^{-1}(Y)), & Y\in\operatorname{img}F\\
0, & \text{otherwise}.
\end{cases}
\]
Since $F$ is essentially injective, this is well defined up to natural
isomorphisms using the axiom of choice. Let $f:Y\rightarrow Z$ be
a morphism in $\mathbf{D}$. If $Y$ do not belong to the image of
$F$, define $\beta(f):\beta(Y)\rightarrow\beta(Z)$ as the unique
map $0\rightarrow\beta(Z)$. In a similar way define $\beta(f)$ when
$Z$ (or both $Y$ and $Z$) do not belong to $F(\mathbf{C})$. Finally,
notice that since $F$ is essentially injective and faithful, it is
an equivalence over its image, so that if both $Y,Z$ belong to $F(\mathbf{C})$,
then to any $f:Y\rightarrow Z$ corresponds a unique $F^{-1}(f)$.
In this case, define $\beta(f)=G(F^{-1}(f))$. It is straightforward
to check the functorial properties of $G$. By definition, $\beta$
is essentially injective when restricted to $F(\mathbf{C})$ and constant
(equal to the null object) in the remaining part. Thus, both $\alpha$
and $\beta$ are qei. Furthermore, by construction the diagram (\ref{quasi_essentially_injective})
commutes up to isomorphisms, giving the desired conjugation and completing
the proof.
\end{svmultproof}

We say that a monoidal category $(\mathbf{A},\otimes,1)$ is $\tau$\emph{-faithful,}
where $\tau:\mathbf{FinSet}\rightarrow\mathbf{A}$ is a given functor,
if the induced monoidal product in $1/\mathbf{A}$ is faithful in
both variables when restricted to the image of $1/\tau$, i.e, if
$f\otimes g=f'\otimes g'$ implies $f=f'$ and $g=g'$ for every $f,g,f',g'\in1/\tau(\mathbf{FinSet})$.
\begin{proposition}
\label{faithful_Feynman_functor} Under the same notations of Proposition
\ref{existence_complete_feynman_functors}, suppose that $(\mathbf{A},\otimes,1)$
is $\tau$-faithful and $\tau$ is $1/$decomposable. In this case,
if $Z:C^{b}\rightarrow(\mathbf{A},\tau)$ is the Feynman rule induced
by a strongly complete Feynman rule $FR$ fulfilling condition \emph{c2)
}or condition \emph{c3)}, then it is qeic to any other complete Feynman
functor $Z':C^{b}\rightarrow(\mathbf{A},\tau)$.
\end{proposition}
\begin{svmultproof}
Since $Z$ is a complete Feynman functor, $Z_{V}$ is essentially
injective for every $V$. Therefore, by the previous lemma it is enough
to prove that Feynman functors induced by strongly complete Feynman
rules taking values into a $\tau$-faithful are also faithful, i.e,
are such that if $Z_{V}(f)=Z_{V}(g)$, then $f=g$. In order to do
this we need to work case by case of Proposition \ref{existence_complete_feynman_functors}.
We will give the proof only for case c3). Case c2) is analogous, needing
only some care with the parentheses. By definition, $Z_{V}(f)$ is
a tensor product between $Z_{V}^{i}(f)$ and $Z_{V,j}^{i}(f)$, with
$1<j\leq b$ and $0\leq i\leq n$. In turn, $Z_{V}^{i}(f)$ is the
map between $\otimes_{v}FR^{i}(v)$ and $\otimes_{w}FR'^{i}(w)$,
with $v\in VG^{i}$ and $w\in VG'^{i}$, given by the restriction
of $\otimes_{v}r^{i}f$. Furthermore, $Z_{V,j}^{i}(f)$ is obtained
in a similar way as a restriction of $\otimes_{e}r_{j}^{i}f$ with
$e\in EG_{j}^{i}$. Therefore, $Z_{V}(f)$ is the restriction of $\otimes_{i,j}\otimes_{v,e}r^{i}f\otimes r_{j}^{i}f$
to an object depending only of $G$ (and not of $f$). So, if $Z_{V}(f)=Z_{V}(g)$,
then 
\[
\otimes_{i,j}\otimes_{v,e}r^{i}f\otimes r_{j}^{i}f=\otimes_{i,j}\otimes_{v,e}r^{i}g\otimes r_{j}^{i}g.
\]
Since both sides are morphisms between $1/\tau VG$ and $1/\tau VG'$,
the fact that $\mathbf{A}$ is $\tau$-faithful implies $r^{i}f=r^{i}g$
and $r_{j}^{i}f=r_{j}^{i}g$. Finally, because $FR$ is strongly proper,
each $FR^{i}$ and each $FR_{j}^{i}$ is an equivalence over their
images. Thus, commutativity of diagrams (\ref{functorial_feynman_rule})
gives us $f^{i}=g^{i}$ and $f_{j}^{i}=g_{j}^{i}$, i.e, $f=g$.
\end{svmultproof}

Putting together all parts of our construction we have the following
general existence and uniqueness theorem.
\begin{theorem}[existence and uniqueness]
Let $C$ be a proper, $(\mathbf{A},\otimes,\cdot^{\vee})$ be a $\tau$-faithful
pseudocontext with null objects, where $\tau:\mathbf{FinSet}\rightarrow\mathbf{A}$
is a $1/$decomposable functor. In this case:
\begin{enumerate}
\item up to normalization, change of coefficients and qeic, there exists
a unique complete Feynman functor $Z:C\rightarrow(\mathbf{A},\tau)$;
\item if $C$ is concrete proper and becomes endowed with a normal structure
$\mathfrak{d}$, then up to change of coefficients and qeic, there
exists a unique monoidal and complete Feynman functor $Z:C\rightarrow(\mathbf{A},\tau)$.
\end{enumerate}
\end{theorem}

\section{Superposition \label{sec_superposition}}

Let $(\mathbf{C},\otimes,1)$ be a monoidal category. We say that
an object $X\in\mathbf{C}$ is the \emph{superposition} of a family
of objects $A_{i}\in\mathbf{C}$, with $1\leq i\leq n$, if there
exists an isomorphism 
\begin{equation}
(...((A_{1}\otimes A_{2})\otimes A_{3})...)\otimes A_{n}\simeq X.\label{superposition}
\end{equation}
The number $n$ is called the \emph{order} of the superposition.
\begin{example}[trivial superposition]
 Each object $X$ admits a superposition of arbitrary order. Indeed,
for given $n$, just take $A_{1}=X$ and $A_{i>1}=1$.
\end{example}
Let $\mathcal{C}\subset\operatorname{Ob}(\mathbf{C})$ be a collection
of objects in $\mathbf{C}$. A \emph{superposition principle} in $\mathcal{C}$
is given by
\begin{enumerate}
\item a bounded from above function $n:\mathcal{C}\rightarrow\mathbb{N}$,
which is equivalent to saying that $n(\mathcal{C})\subset\mathbb{N}$
is finite\footnote{Indeed, due to the well-ordering principle a subset $S\subset\mathbb{N}$
is always bounded from below, hence it is bounded from above iff it
is bounded. Furthermore, any infinite subset of $\mathbb{N}$ is in
bijection with $\mathbb{N}$, so that it is unbounded. Finally, every
finite subset of $\mathbb{N}$ is bounded.}. The maximum of $n(\mathcal{C})$ will be denoted by $N$;
\item a function $\mathcal{S}:\mathcal{C}\rightarrow\operatorname{Ob}(\mathbf{C})$
assigning to each object $X$ in $\mathcal{C}$ a family of objects
$\mathcal{S}X_{i}$ in $\mathbf{C}$, with $1\leq i\leq N$, such
that
\begin{enumerate}
\item for $i>n(X)$ we have $SX_{i}\simeq1$;
\item after taking the tensor product between the $SX_{i}$, in the same
order that in (\ref{superposition}) we get a superposition for $X$.
\end{enumerate}
\end{enumerate}
We say that a superposition principle is \emph{functorial} if $\mathcal{C}$
is actually a (non necessarily monoidal) subcategory of $\mathbf{C}$
such that the function $\mathcal{S}:\mathcal{C}\rightarrow\operatorname{Ob}(\mathbf{C})$
extends to a functor $\mathcal{S}:\mathcal{C}\rightarrow\mathbf{C}$.
\begin{example}[trivial superposition principle]
 In any set $\mathcal{C}$ we have a trivial superpostion principle,
which assigns to each $X\in\mathcal{C}$ its corresponding trivial
superposition.
\end{example}
In the following we will show that nontrivial Feynman functors behave
as a bridge between reconstruction conjectures and nontrivial superposition
principles. In order to do this, let $\mathcal{D}$ be a deleting
process, which assigns to each applicable prestack $C\in\mathfrak{C}$
its prestack $\mathcal{D}C$ of pieces, and let us define a \emph{structure
of disjoint pieces }in some $C\in\mathfrak{C}$ as:
\begin{enumerate}
\item a rule that for each $V$ associates a decomposition $V=\coprod_{i}\overline{V}_{i}$;
\item a subprestack $K\subset C$ (not necessarily applicable);
\item for each $V$ a functor $\kappa_{V}:\mathcal{D}C_{V}\rightarrow\prod_{i}K_{\overline{V}_{i}}$.
\end{enumerate}
\begin{example}[representable reconstruction]
 If each $\mathcal{D}C_{V}$ is representable, then it maps colimits
into limits, so that for any decomposition $V=\coprod_{i}\overline{V}_{i}$
we have $\mathcal{D}C_{\coprod_{i}\overline{V}_{i}}\simeq\prod_{i}\mathcal{D}C_{\overline{V}_{i}}.$
So, if $DC_{V}\subset C_{V}$, we can take $K_{\overline{V}_{i}}=\mathcal{D}C_{\overline{V}_{i}}$
and $\kappa_{V}$ as the previous isomorphism.
\begin{example}[disjoint reconstruction]
 The disjoint context $\mathcal{D}C_{V}=\sqcap C_{V}$ has a canonical
structure of disjoint pieces with $\overline{V}_{i}=V-v_{i}$, $K=DC$
and $\kappa=id$.
\end{example}
\end{example}
\begin{theorem}
\label{induced_superposition} Let $(C,\xi)$ be a proper prestack
of hypergraphs which is an applicable prestack of certain reconstruction
context $\mathcal{D}$. Suppose given a structure of disjoint pieces
in $C$. Then each nontrivial strong monoidal Feynman functor $Z:C\rightarrow(\mathbf{A},\tau)$
induces a nontrivial superposition principle in a set $\mathcal{A}_{\tau}$
of analytic expressions in $\mathbf{A}$, which becomes endowed with
canonical map $\lambda_{Z}:\mathcal{G}\rightarrow\mathcal{A}_{\tau}$,
where $\mathcal{G}$ is the set of all isomorphism classes of hypergraphs
$G\in C_{V}$ for every $V$.
\end{theorem}
\begin{svmultproof}
We will give the proof for the case in which $(\mathbf{A},\otimes,1)$
is strict. The general case is analogous, only needing some care with
the parentheses. For each $V$, we have the functor $\kappa_{V}:\mathcal{D}C_{V}\rightarrow\prod_{i}K_{\overline{V}_{i}}$.
On the other hand, since $C$ is proper for any $W,W'$ we have $\xi:C_{W}\times C_{W'}\rightarrow C_{W\sqcup W'}$,
which clearly extends to a functor $\xi:\prod_{i}C_{W_{i}}\rightarrow C_{\coprod_{i}W_{i}}$,
where $W_{i}$ is any finite family. Because $K\subset C$, such functors
can be restricted to $K$, giving $\xi:\prod_{i}K_{W_{i}}\rightarrow K_{\coprod_{i}W_{i}}$.
Taking $W_{i}=\overline{V}_{i}$, let us consider $\xi$ restricted
to the image of $\kappa_{V}$. Since $Z$ is strong monoidal, for
every $\mathcal{D}G\in\mathcal{D}C_{V}$ we have an isomorphism 
\begin{equation}
Z_{\coprod_{i}\overline{V}_{i}=V}(\xi(\mathcal{D}G))\simeq\otimes_{i}Z_{\overline{V}_{i}}(\mathcal{D}G_{i}),\label{theorem_superposition}
\end{equation}
where $\mathcal{D}G_{i}\in K_{\overline{V}_{i}}$ are the components
of $\kappa_{V}(\mathcal{D}G)$. We can regard the isomorphism above
as superposition principle for $Z(\xi(\mathcal{D}G))$. Recall that
for every $V$ and every $\mathcal{D}G\in\mathcal{D}C_{V}$, $Z_{V}(\mathcal{D}G)\in\mathbf{A}_{\tau V\mathcal{D}G}$.
Therefore, if $\mathcal{A}_{\tau}\subset\operatorname{Ob}(1/\mathbf{A})$
is the set of all analytic expressions in $\tau V$, for every $V$,
then $Z_{V}(\mathcal{D}G)\in\mathcal{A}_{\tau}$ for each $V$ and
each $\mathcal{D}G$. This means that when varying $V$ and $\mathcal{D}G$
in (\ref{theorem_superposition}) we get a superposition principle
in $\mathcal{A}_{\tau}$. Let $\mathcal{DG}$ be the set of all isomorphism
classes of $\mathcal{D}G\in\mathcal{D}C_{V}$ for every $V$. So we
have an obvious function $\Lambda_{Z}:\mathcal{DG}\rightarrow\mathcal{A}_{\tau}$
assigning $Z_{V}(\mathcal{D}G)$ to each $\mathcal{D}G$. Finally,
recall that in a reconstruction context there exists a transformation
$\gamma_{V}:C_{V}\rightarrow DC_{V}$, which produces $\gamma:\mathcal{G}\rightarrow\mathcal{DG}$.
By composing with $\Lambda_{Z}$ we get the desired $\lambda_{Z}:\mathcal{G}\rightarrow\mathcal{A}_{\tau}$.
\end{svmultproof}

\begin{corollary}
\label{recontruction_iff_lambda_injective} Let $C\in\mathfrak{C}$
be an applicable prestack of a reconstruction context $\mathcal{D}$.
Suppose that $C$ is proper and endowed with a structure of disjoint
pieces. Then $\mathcal{D}$-RC-$C$ holds only if for every nontrivial
complete monoidal Feynman rule $Z:C\rightarrow(\mathbf{A},\tau)$
the corresponding map $\lambda_{Z}$ is injective.
\end{corollary}
\begin{svmultproof}
Assume $\mathcal{D}$-RC-$C$ holds, which means that $\gamma_{V}:C_{V}\rightarrow DG_{V}$
is essentially injective for every $V$, and let $Z:C\rightarrow(\mathbf{A},\tau)$
be a complete monoidal Feynman rule. Recall that $\lambda_{Z}:\mathcal{G}\rightarrow\mathcal{A}_{\tau}$
is the composition between $\gamma:\mathcal{G}\rightarrow\mathcal{DG}$
and $\Lambda_{Z}:\mathcal{DG}\rightarrow\mathcal{A}_{\tau}$. Completeness
of $Z$ implies $\Lambda_{Z}$ injective and essential injectivity
of $\gamma_{V}$ makes $\gamma$ injective, so that $\lambda_{Z}$
is also injective. Reciprocally, suppose that $\lambda_{Z}$ is injective
for some complete monoidal Feynman rule. By definition, $\lambda_{Z}=\Lambda_{Z}\circ\gamma$.
Since injective functions are monomorphisms, $\gamma_{Z}$ injective
implies $\gamma$ injective, which is equivalent to saying that $\gamma_{V}$
is essentially injective for every $V$, so that $\mathcal{D}$-RC-$C$
holds.
\end{svmultproof}

\begin{remark}
As a consequence of Proposition \ref{prop_morphism_conjecture} and
Proposition \ref{prop_implication}, the previous corollary can be
improve by replacing the hypothesis of $\mathcal{D}$-RC-$C$ holding
with the existence of left/right morphisms $F:\mathcal{D}\rightarrow\mathcal{D}'$
or left/right $C$-implications to some other reconstruction context.
\end{remark}
\begin{corollary}
Let $C\subset S_{\mathfrak{s}}^{b,\ell}$ be some concretely proper
prestack of labeled structured bounded hypergraphs, regarded as an
applicable prestack for the standard reconstruction context $\mathcal{D}=d$
and endowed with the canonical structure of disjoint pieces. Then
for every complete monoidal Feynman functor $Z:C\rightarrow(\mathbf{A},\tau)$
the induced map $\lambda_{Z}$ in injective. Explicitly, 
\[
Z_{V}(G)\simeq Z_{V}(G')\quad\textit{iff}\quad\bigotimes_{v\in V}Z(G-v)\simeq\bigotimes_{v\in V}Z(G'-v).
\]
\end{corollary}
\begin{svmultproof}
Immediate from Corollary \ref{reconstruction_labeled_structured},
Corollary \ref{recontruction_iff_lambda_injective} and from the definition
of $\lambda_{Z}$.
\end{svmultproof}

\section{Applications \label{applications}}

As the first application we present an existence result for representations
of hypergraph categories in monoidal categories.
\begin{proposition}
\label{application_1} Let $C^{b}$ be $b$-bounded prestack of hypergraphs.
Let $(\mathbf{A},\otimes,\cdot^{\vee})$ be a pseudocontext which
is $\tau$-faithful relative to a $1/$decomposable functor $\tau:\mathbf{FinSet}\rightarrow\mathbf{A}$.
Suppose one of the following conditions:
\begin{enumerate}
\item[c1)] $C^{b}\subset S_{\mathfrak{s}}^{b,\ell}$ for some functor of structures
$\mathfrak{s}$;
\item[c2)] $(\mathbf{A},\otimes,1)$ is a strict monoidal category.
\end{enumerate}
In this case, the choice of a normal structure $\mathfrak{d}$ for
$C^{b}$ induces an equivalence between the category $C_{V}^{b}$,
for each $V$, and some subcategory of $1/\mathbf{A}$.

\end{proposition}
\begin{svmultproof}
Notice that we are in the hypothesis of Proposition \ref{faithful_Feynman_functor}
and from its proof we see that if there exists some strongly complete
Feynman rule $FR$ in $(C^{b},\mathfrak{d})$, then it induces a Feynman
functor $Z:C^{b}\rightarrow(\mathbf{A},\tau)$ such that each $Z_{V}:C_{V}^{b}\rightarrow1/\mathbf{A}$
is essentially injective and faithful, so that they are equivalences
over their images. Since $\mathfrak{d}$ is chosen normal, it follows
from Corollary \ref{corollary_complete_functors} that these Feynman
rules really exist.
\end{svmultproof}

\subsection{Mapping Class Group and Ribbon Graphs}

One of the main problems of manifold topology is to determine the
mapping class group $\operatorname{MCG}(M)$ of a given manifold $M$,
i.e, the quotient of the diffeomorphism group $\operatorname{Diff}(M)$
by the path-component at the identity. It is a remarkable fact that
for marked surfaces that object can be described by the category of
ribbon graphs. More precisely, if $\mathbf{Ribb}_{3}^{con}\subset\mathbf{Ribb}$
denotes the category of connected ribbon graphs whose vertices are
at least trivalent and with morphisms given by ribbon graphs isomorphisms,
then there exists a homotopy equivalence 
\[
\vert\mathbf{Ribb}_{3}^{con}\vert\simeq\coprod_{[\Sigma]\in\operatorname{Iso}(\mathbf{Diff}_{2}^{*})}B\operatorname{MCG}(\Sigma),
\]
where $\vert\mathbf{C}\vert$ and $BG$ denotes, respectively, the
geometric realization of a category and the classifying space of a
topological group. Furthermore, the coproduct is taken over the diffeomorphic
classes of all marked surfaces, except for two exceptional cases:
the sphere $\mathbb{S}^{2}$ with one and with two marked points.

This result was proven using different methods in \citep{ribbon_graphs_1,ribbon_graphs_2,ribbon_graphs_3,ribbon_3,ribbon_4}.
As a second application of the existence of complete Feynman rules
we give an independent proof of a related result. Indeed, we will
show that for any fixed vertex set $V$, the category $\mathbf{Ribb}_{V}$
can be regarded as a subcategory $\mathbf{A}_{V}$ of 
\[
\coprod_{[\Sigma]\in\operatorname{Iso}(\sqcup\mathbf{Diff}_{2}^{*})}\mathbf{B}\operatorname{MCG}(\Sigma),
\]
where $\mathbf{B}G$ is the delooping groupoid of a group and $\sqcup\mathbf{Diff}_{2}^{*}$
means that the coproduct is take over arbtirary finite coproducts
of marked surfaces.

Let $\mathbf{Diff}^{*}$ be the category of marked manifolds, i.e,
pairs $(M,S)$, where $S\subset M$ is some finite subset of mutually
distinct points, and morphisms given by smooth maps $f:M\rightarrow M'$
such that $f(S)\subset S'$. The category has coproducts given by
$\coprod_{i}(M_{i},S_{i})=(\coprod_{i}M_{i},\coprod_{i}S_{i})$. Two
marked manifolds are isomorphic only if $S\simeq S'$, which means
that in the isomorphism classes only the number of marked points matter.
We will write $M^{s}$ instead of $(M,S)$, where $s=\vert S\vert$.
Let $\mathbf{Sing}$ be the proper class of singletons and fix an
injective class function $\omega:\operatorname{Iso}(\mathbf{Diff}^{*})\rightarrow\mathbf{Sing}$.
For instance, we could take $\omega([M])=\{[M]\}$. Define a category
$\operatorname{MCG}_{\mathbf{Diff}^{*}}^{\omega}$ as follows. Objects
are given by the image of $\omega$, i.e, we have an object for each
isomorphism class of marked manifolds and this object which is a singleton
$\omega([M^{s}])=*_{[M^{s}]}$. Furthermore, there are morphisms between
$*_{[M^{s}]}$ and $*_{[N^{r}]}$ iff $[M^{s}]=[N^{r}]$ and in this
case $\mathrm{Mor}(*_{[M^{s}]},*_{[M^{s}]})=\operatorname{MCG}(M^{s})$.
In particular, $(\operatorname{MCG}_{\mathbf{Diff}^{*}}^{\omega})^{op}=\operatorname{MCG}_{\mathbf{Diff}^{*}}^{\omega}$.
Furthermore, since 
\[
\operatorname{MCG}_{\mathbf{Diff}^{*}}^{\omega}\simeq\coprod_{[M^{s}]\in\operatorname{Iso}(\mathbf{Diff}^{*})}\mathbf{B}\operatorname{MCG}(M^{s}),
\]
the left-hand side does not depends of $\omega$. Coproducts pass
to $\operatorname{MCG}_{\mathbf{Diff}^{*}}^{\omega}$ by taking $*_{[M^{s}]}\sqcup*_{[N^{r}]}=*_{[M^{s}\sqcup N^{r}]}$.
Furthermore, if $f\in\operatorname{MCG}(M^{s})$ and $g\in\operatorname{MCG}(N^{r})$,
define $f\sqcup g:*_{[M\sqcup N]}\rightarrow*_{[M\sqcup N]}$ as the
coproduct in $\mathbf{Diff}^{*}$.

Consider a full subcategory $\mathbf{A}\subset\operatorname{MCG}_{\mathbf{Diff}^{*}}^{\omega}$
closed by finite coproducts, so that we can take the cocartesian monoidal
structure and since $\mathbf{A}^{op}=\mathbf{A}$, we get a pseudocontext
structure $(\mathbf{A},\sqcup,\cdot^{\vee})$ with $\cdot^{\vee}:\mathbf{A}\rightarrow\mathbf{A}$
some endofunctor of $\mathbf{A}$. These functors are in 1-1 correspondence
with rules assigning manifolds $M^{s}$ such that $\omega([M^{s}])\in\mathbf{A}$
to manifolds $N$ such that $\omega([N^{r}])\in\mathbf{A}$, together
with a group homomorphism $\operatorname{MCG}(M^{s})\rightarrow\operatorname{MCG}(N^{r})$.
For instance, if in $\mathbf{A}$ there exists some manifold $X^{o}$
whose mapping class group is completely understood, we can take $\cdot^{\vee}$
as some functor constant in such a manifold, which is equivalent to
giving a representation of every $\operatorname{MCG}(M^{s})$ in $\operatorname{MCG}(X^{o})$.
But we can also simply take $\cdot^{\vee}=id_{\mathbf{A}}$. Let $S\neq\varnothing$
be a connected manifold such that $\omega([S^{s}])\in\mathbf{A}$
for every $s$ and define $\tau_{S}:\mathbf{FinSet}\rightarrow\mathbf{A}$
as follows. For each finite set $X$ we take $\tau_{S}(X)=\omega([S^{\vert X\vert}])$.
Furthermore, if $f:X\rightarrow Y$ is a map between finite sets,
then $\tau_{S}(f)$ is nontrivial iff $X\simeq Y$ and in this case
$\tau_{S}(f)=id_{S^{\vert X\vert}}$. objects and

Now, recall that each construction in this article was made in $1/\mathbf{A}$.
The only reason for doing this was to have a frame closer to physics
interpretation. Indeed, in this way we can talk about analytic expressions
itself, while when working in $\mathbf{A}$ we can only talk about
the object of all analytic expressions. Even so, all definitions,
statements and demonstrations work \emph{ipsi literis} in $\mathbf{A}$.
With this in mind we can search for Feynman functors taking values
in $(\mathbf{A},\tau_{S})$ for the pseudocontext $(\mathbf{A},\sqcup,\cdot^{\vee})$
defined above.
\begin{proposition}
\label{MCG} For each connected non-empty manifold $S$ such that
$\omega(S^{s})\in\mathbf{A}$ for every $s$, the pseudocontext $\mathbf{A}\subset\operatorname{MCG}_{\mathbf{Diff}^{*}}^{\omega}$
is $\tau_{S}$-faithful and $\tau_{S}$ is decomposable.
\end{proposition}
\begin{svmultproof}
Notice that $\tau_{S}(\mathbf{FinSet})$ is the category whose objects
are $\omega([S^{s}])$, with $s\geq1$, and whose only morphisms are
the identities $id_{S^{s}}$, for $s\geq1$. Therefore, $(\mathbf{A},\sqcup,\varnothing)$
is clearly $\tau_{S}$-faithful. In order to see that $\tau_{S}$
is decomposable, given finite sets $X$ and $X'$, suppose we have
decompositions 
\[
\tau_{S}(X)\simeq\omega([M^{s}])\sqcup\omega([N^{r}])\quad\text{and}\quad\tau_{S}(X')\simeq\omega([M'^{s'}])\sqcup\omega([N'^{r'}]).
\]
This implies $S^{\vert X\vert}\simeq M^{s}\sqcup N^{r}$ and $S^{\vert X'\vert}\simeq M'^{s'}\sqcup N'^{r'}$.
Since $S$ is connected, either we have the following configuration
or we have one of the other seven permutations:
\begin{itemize}
\item $M^{s}\simeq S^{\vert X\vert}$ and $N^{r}=\varnothing$, together
with $M'^{s'}\simeq S^{\vert X'\vert}$ and $N'^{r'}=\varnothing$.
\end{itemize}
We should proceed case by case, but since everything is analogous
we will work only with the above configuration. Let $f:X\rightarrow X'$
be an isomorphism. Then $\vert X\vert=\vert X'\vert$, implying $M^{s}=M'^{s'}\simeq S^{\vert X\vert}$.
Since $\tau_{S}(f)=id_{S^{\vert X\vert}}$, we just take $\alpha=id_{M^{s}}$.
Furthermore, since $N^{r}=\varnothing=N'^{r'}$, there exists a unique
$\beta:N^{r}\simeq N'^{r'}$ and we clearly have $\tau_{S}(f)=\alpha\sqcup\beta$.
\end{svmultproof}

\begin{corollary}
Let $C^{b}$ be a $b$-bounded prestack of hypergraphs. For each normal
structure $\mathfrak{d}$ and each non-empty connected manifold $S$
there exists an equivalence between $C_{V}^{b}$, for every $V$,
and a subcategory $\mathbf{A}_{V}$ of any pseudocontext $\mathbf{A}\subset\operatorname{MCG}_{\mathbf{Diff}^{*}}^{\omega}$
such that $\omega(S^{s})\in\mathbf{A}$ for every $s$.
\end{corollary}
\begin{svmultproof}
Straighforward from Proposition \ref{MCG} and Proposition \ref{application_1}.
\end{svmultproof}

Let $\mathbf{\operatorname{MCG}^{\omega}}_{\sqcup\mathbf{Diff}_{2}^{*}}\subset\operatorname{MCG}_{\mathbf{Diff}^{*}}^{\omega}$
denote the subcategory generated by compact orientable marked surfaces
and finite disjoint unions of them.
\begin{corollary}
The choice of a normal decomposition $\mathfrak{d}$ induces an equivalence
between the category $\mathbf{Ribb}_{V}$ and some subcategory of
$\mathbf{\operatorname{MCG}^{\omega}}_{\sqcup\mathbf{Diff}_{2}^{*}}$.
\end{corollary}
\begin{svmultproof}
Just recall that any prestack of graphs is 2-bounded and then apply
the last corollary.
\end{svmultproof}

\begin{remark}
Our construction, however, does not give much information about the
homotopy type of the classifying spaces $B\operatorname{MCG}(M)$.
Indeed, since we included the empty manifold in $\operatorname{MCG}_{\mathbf{Diff}^{*}}^{\omega}$
and the empty graph in each $C_{V}^{b}$, both admit initial objects,
so that their geometric realizations are automatically contractible.
\end{remark}

\subsection{Perturbative QFT}

As a final application, let us show that the validity of a reconstruction
conjecture induces a new superposition principle in perturbative QFT.
A \emph{classical field theory }$\mathcal{S}$ (following \citep{Costello})
is given by the following data:
\begin{enumerate}
\item a vector bundle $\pi:E\rightarrow M$ over a compact riemannian manifold
$M$;
\item a positive generalized laplacian $\mathcal{D}$ in $E^{!}=E^{\vee}\otimes\mathrm{Dens}_{M}^{s,+}\simeq\mathrm{Hom}(E;\mathrm{Dens}_{M}^{s,+})$,
from which we build the free functional $S_{0}[s]=\int_{M}\mathcal{D}s$.
Typical examples occur when $\mathcal{D}s=\langle s,Ds\rangle$, where
$D$ is a generalized laplacian in $E$ and $\langle\cdot,\cdot\rangle:E\otimes E\rightarrow\mathrm{Dens}_{M}^{s,+}$
is a symmetric bundle map;
\item a differential operator $\mathfrak{D}$ between $E^{!}$ and $E$,
which is formally symmetric, i.e, $\mathfrak{D}^{!}=\mathfrak{D}$
and such that $\mathcal{D}\circ\mathfrak{D}^{!}=\mathfrak{D}\circ\mathcal{D}^{!}$;
\item an element $I\in\Gamma(E)[[\hbar]]$, giving the full action functional
$S=S_{0}+I$.
\end{enumerate}
Since $\Gamma(E)$ is a nuclear Fr\'{e}chet space, it belongs to
context $(\mathbf{NucFrec},\otimes_{p},\cdot^{\vee})$ of Example
\ref{ex_nuc_frechet}. Let us take $\tau:\mathbf{FinSet\rightarrow\mathbf{NucFrec}}$
as constant in $\Gamma(E)$. From any classical field theory (in the
above sense) we can extract a full subcategory $\mathbf{Feyn}(\mathcal{S})\subset\mathbf{Feyn}$
simply by doing the standard Feynman graph expansion of an action
funcional \citep{Costello}. Varying $V$ on $\mathbf{Feyn}(\mathcal{S})_{V}$
we get a prestack $C_{\mathcal{S}}$. Let us call the pair $(\mathcal{S},C_{\mathcal{S}})$
the \emph{pertubative QFT} of $\mathcal{S}$. Notice, on the other
hand, that from data 1-3 above we can extract a tensor $P\in\operatorname{Sym}^{2}\Gamma(E)$,
obtained as follows. Since $M$ is compact, $\mathcal{D}$ has a smooth
heat kernel $K_{t}\in\Gamma(E^{!})\otimes\Gamma(E)\otimes C^{\infty}(\mathbb{R}_{\geq0})$.
Composing with $\mathfrak{D}$ we get an element of $\mathcal{K}_{t}\in\Gamma(E)\otimes\Gamma(E)\otimes C^{\infty}(\mathbb{R}_{\geq0})$.
Because $\mathfrak{D}$ is symmetric, $\mathcal{K}_{t}$ is symmetric
too. By means of integrating we get $P\in\mathrm{Sym}^{2}(\Gamma(E))$.
We can then get a Feynman rule $FR:(C_{\mathcal{S}},\mathfrak{d})\rightarrow(\mathbf{NucFrec},\tau)$
from Example \ref{physics_feynman_rule} by fixing $FR_{2}^{1}$ as
constant in $P$ and $FR^{1}$ as determined by $I$. We also take
$FR^{0}$ constant due to the indistinguishability of quantum particles.
Since $C_{\mathcal{S}}$ is concretely proper, it follows from Example
\ref{concret_proper} and Proposition \ref{oplax_feynman_rule_functor}
that the induced Feynman functor $Z_{\mathcal{S}}:C_{\mathcal{S}}\rightarrow(\mathbf{NucFrec},\tau)$
is oplax monoidal. Let us define a \emph{superposition principle}
for $(\mathcal{S},C_{\mathcal{S}})$ as a superposition principle
in the image of $Z_{\mathcal{S}}$.
\begin{proposition}
Let $(\mathcal{S},C_{\mathcal{S}})$ be a perturbative QFT whose Feynman
functor $Z_{\mathcal{S}}$ is strong monoidal. If $C_{\mathcal{S}}$
is an applicable prestack for a reconstruction context $\mathcal{D}$
we have a nontrivial superposition principle for $(\mathcal{S},C_{\mathcal{S}})$.
\end{proposition}
\begin{svmultproof}
Just apply Theorem \ref{induced_superposition}.
\end{svmultproof}

\begin{acknowledgements}
Y. X. Martins was supported by CAPES. Both authors would like to thank
Bhalchandra Digambar Thatte for stimulating and helpful discussions
on the reconstruction conjecture.
\end{acknowledgements}

\end{document}